\documentclass[aps,prd,twocolumn,showpacs,floatfix,preprintnumbers,amsmath,amssymb,nofootinbib]{revtex4-1}

\input epsf
\usepackage{graphicx}
\usepackage{color}

\newcommand{\beq}{\begin{equation}}
\newcommand{\eeq}{\end{equation}}
\newcommand{\barr}{\begin{eqnarray}}
\newcommand{\earr}{\end{eqnarray}}

\newcommand{\rme}{\textrm{e}}

\newcommand{\Ly}{\textrm{Ly}}

\newcommand{\Tm}{T_{\rm m}}
\newcommand{\Tr}{T_{\rm r}}
\newcommand{\nh}{n_{\rm H}}

\newcommand{\apjl}{Astrophys. J. Lett.}
\newcommand{\aap}{Astron. Astrophys.}
\newcommand{\apjs}{Astrophys. J. Suppl. Ser.}

\newcommand{\mnras}{Mon. Not. R. Astron. Soc.}

\newcommand{\memras}{Mem. R. Astr. Soc. }

\newcommand{\lsim}{\mathrel{\hbox{\rlap{\lower.55ex\hbox{$\sim$}} \kern-.3em \raise.4ex \hbox{$<$}}}}
\newcommand{\gsim}{\mathrel{\hbox{\rlap{\lower.55ex\hbox{$\sim$}} \kern-.3em \raise.4ex \hbox{$>$}}}}
\begin{document}
\title{\textsc{HyRec}: A fast and highly accurate primordial hydrogen and helium recombination code}

\author{Yacine Ali-Ha\"imoud} 
\author{Christopher M. Hirata} 
\affiliation{California Institute of Technology, Mail Code 350-17, Pasadena, CA 91125}

\date{\today}
\begin{abstract}
We present a state-of-the-art primordial recombination code, \textsc{HyRec}, including all the physical effects that have been shown to significantly affect recombination. The computation of helium recombination includes simple analytic treatments of hydrogen continuum opacity in the He I $2^1P^o - 1^1S$ line, the He I] $2^3P^o-1^1S$ line, and treats feedback between these lines within the on-the-spot approximation. Hydrogen recombination is computed using the effective multilevel atom method, virtually accounting for an infinite number of excited states. We account for two-photon transitions from $2s$ and higher levels as well as frequency diffusion in Lyman-$\alpha$ with a full radiative transfer calculation. We present a new method to evolve the radiation field simultaneously with the level populations and the free electron fraction. These computations are sped up by taking advantage of the particular sparseness pattern of the equations describing the radiative transfer. The computation time for a full recombination history is $\sim 2$ seconds. This makes our code well suited for inclusion in Monte Carlo Markov chains for cosmological parameter estimation from upcoming high-precision cosmic microwave background anisotropy measurements.
\end{abstract}

\maketitle
\section{Introduction}

Until recently, primordial recombination was considered one of the few solved problems in astrophysics and cosmology. The seminal works of Peebles \cite{Peebles} and Zeldovich \emph{et al.} \cite{Zeldovich_et_al} in the 1960s established that hydrogen recombination did not proceed in Saha equilibrium, and that two-photon decays from $2s$ are critical to the recombination dynamics because of the very low escape rate of Lyman-$\alpha$ photons. They provided a simple effective three-level atom model to compute primordial hydrogen recombination histories. With the advent of high-precision cosmic microwave background (CMB) experiments such as \emph{WMAP} \cite{WMAP7} and \emph{Planck} \cite{Planck}, it has become clear that these early calculations are not sufficently accurate for an unbiased estimate of cosmological parameters \cite{Hu1995, Lewis06, Chluba10_uncertainties}. Uncertainties in the recombination history indeed propagate to the visibility function and ultimately to the predicted CMB temperature and polarization anisotropies. 

These considerations have motivated Seager \emph{et al.} \cite{Recfast_long, Recfast_short} to extend Peebles' effective three-level atom model to a multilevel atom calculation. They showed that accounting for excited states of hydrogen leads to a speed up of recombination at late times. Their commonly used recombination code \textsc{RecFast} approximately reproduces these results by solving an effective three-level atom model with an artificially enhanced recombination coefficient. While this model is sufficiently accurate for current CMB data analysis, it does not meet the $\sim 0.1\%$ accuracy target for \emph{Planck}.

Several physical effects have since then been shown to significantly affect hydrogen and helium recombination. First, the multilevel computations of Seager \emph{et al.} assumed statistical equilibrium among the angular momentum substates of a given energy shell of hydrogen. At late times, this assumption breaks down, and an accurate multilevel atom computation should resolve the angular momentum substates \cite{RMCS06, CRMS07}. Whereas it is a straightforward conceptual generalization of previous works, this problem can be computationally challenging. Several works have tackled the ``high-$n$ problem'', including increasingly larger numbers of excited states of hydrogen to reach sufficient accuracy \cite{RMCS06, CRMS07, Grin_Hirata, Chluba_Vasil}. The ``standard'' multilevel atom (MLA) method used in these works requires solving for the abundances of all the excited states at every timestep, which makes the computation very time-consuming. In a recent paper (Ref.~\cite{EMLA}, hereafter ``Paper I''), we have introduced a new method of solution for the multilevel atom problem, which allows to only solve for the populations of a few excited states (typically, $2s$ and the low-lying $p$ states), provided one uses precomputed \emph{effective} bound-bound and bound-free transition rates, which contain all the information about the highly excited states of hydrogen. This method alleviates the computational difficulty associated with the highly excited states of hydrogen and is key to the speed of the recombination code presented here.

Another important aspect of the recombination problem is that of radiative transfer in the vicinity of the Lyman-$\alpha$ line. In its early stages, hydrogen recombination is mostly controlled by the slow escape (via redshifting) of photons from the Lyman-$\alpha$ line and the rate of two-photon decays from the $2s$ state. Accurate values for these rates require treatements of the radiation field that go beyond the simple Sobolev approximation \cite{Recfast_long, AGH10}. Important corrections include feedback from higher-order lines \cite{CS07, Kholupenko_Deuterium}, time-dependent effects in Ly-$\alpha$ \cite{CS09a}, and frequency diffusion due to resonant scattering \cite{Dubrovich_Grachev08, Hirata_Forbes, CS09c}. An accurate $2s-1s$ two-photon decay rate also requires following the radiation field to account for stimulated decays \cite{CS06} and absorption of non-thermal photons \cite{Kholupenko06, Hirata_2photon}. Dubrovich \& Grachev \cite{Dubrovich_Grachev05} suggested that two-photon transitions from higher levels may have a significant effect on the recombination history. Later computations confirmed this idea \cite{CS_2photon}, and provided an accurate treatement of radiative transfer in the presence of two-photon transitions, as well as a solution for the double-counting problem (which arises for resonant two-photon transitions, already included in the one-photon treatment as ``1+1'' transitions) \cite{Hirata_2photon,CS09b}.

The accuracy requirement is less stringent for primordial helium recombination, as it is completed by $z \sim 1700$, much earlier than the peak of the visibility function. Corrections at the percent level are still important, and several works have been devoted to the problem \cite{Matsuda69, Matsuda71, Hu1995, Recfast_long, Dubrovich_Grachev05, Kholupenko_He, Hirata_SwitzerI, Hirata_SwitzerII, Hirata_SwitzerIII, Rubino08_He, Kholupenko08_He}. The most important effect is continuum opacity in the He I 2$^1P^o-1^1S$ line due to photoionization of neutral hydrogen, which requires a detailed line + continuum radiative transfer analysis \cite{Kholupenko_He, Hirata_SwitzerI, Rubino08_He}.  The inclusion of the intercombination line He I] 2$^3P^o-1^1S$ is also significant.

Several other processes have been investigated and shown not to be significant for CMB anisotropies, for example the effects of the isotopes D and $^3$He \cite{Hirata_SwitzerIII, Kholupenko_Deuterium, AGH10}, lithium recombination \cite{SH05}, quadrupole transitions \cite{Grin_Hirata}, high-order Lyman line overlap \cite{AGH10}, and Thomson scattering \cite{CS09c, AGH10}.  Collisional processes are negligible for helium recombination \cite{Hirata_SwitzerIII}; for hydrogen recombination, collisional corrections appear to be small \cite{CVD10}, but whether they are truly negligible is still under investigation.

Previous works have all concentrated on one or a few aspects of the primordial recombination problem. Producing a complete and fast recombination code has so far been hindered by the computational burden previously associated with the high-$n$ problem. Given that this problem is now solved, and that it seems that the main radiative transfer effects have now all been identified, it is timely to deliver a single code that computes an accurate hydrogen and helium recombination history and incorporates all the relevant physics. The purpose of this paper is to introduce our new recombination code, \textsc{HyRec}, which is publicly available\footnote{\textsc{HyRec} is available for download at the following url: http://www.tapir.caltech.edu/$\sim$yacine/hyrec/hyrec.html}, and can compute a highly accurate recombination history (with errors at the level of a few times $10^{-3}$ for helium recombination and a few times $10^{-4}$ for hydrogen recombination) in only $\sim$2 seconds on a standard laptop. Our code does not account for collisional transitions in hydrogen, as their rates are poorly known. When accurate rates are available and if collisional transitions are shown to significantly impact recombination, we will update our code with the appropriate effective rates. 

Recently, a similar work has been carried out by Chluba \& Thomas \cite{Chluba_Thomas}, also relying on the effective MLA method presented in Paper I. The code they present includes the same physics as ours. The main difference is the treatment of radiative transfer. In Ref.~\cite{Chluba_Thomas}, an ``order zero'' recombination history is first computed, with a simple treatment of radiative transfer. The radiative transfer equation is then solved, given this order zero history. Corrections to the net decay rates to the ground state are then evaluated, and used to compute a corrected recombination history. This procedure can in principle be iterated, but because the corrections are small, it is essentially converged after one iteration. Our solution, on the other hand, is non-perturbative, in the sense that we solve \emph{simultaneously} for the radiation field and the recombination history. A detailed code comparison is in progress, and a full error budget will be presented once it is completed.

This paper is organized as follows. In Sec.~\ref{sec:overview}, we review the effective three-level atom model and discuss its limitations. We then review the standard MLA computation and describe the effective MLA method in Sec.~\ref{sec:MLA}. We show that weak transitions to the ground state from excited states with $n \geq 3$ can in fact be accounted for almost exactly with an effective four-level atom model. Two-photon processes and frequency diffusion are formally described in Sec.~\ref{sec:two-photon formal}. In Sec.~\ref{sec:two-photon numerical}, we present our numerical solution for the radiative transfer equation. We use a new method of solution, extending that of Ref.~\cite{Hirata_2photon} to account for frequency diffusion, that allows to solve for the atomic populations and the radiation field simultaneously. We describe our treatment of helium recombination in Sec.~\ref{sec:helium}. We conclude in Sec.~\ref{sec:conclusion}. Appendix \ref{app:effective-rates} demonstrates some relations satified by the effective rates, Appendix \ref{app:extrapolation} describes how we extrapolate the effective rates to an infinite number of excited states, Appendix \ref{app:ODE} describes our ordinary differential equation (ODE) integrator, and Appendix \ref{app:post-Saha} derives an analytic expression for the post-Saha expansion used at early times in hydrogen recombination.

Throughout this paper we use a flat background $\Lambda$CDM cosmology with $T_0 = 2.728$ K, $\Omega_b h^2 = 0.022$, $\Omega_m h^2 = 0.13$, $\Omega_{\Lambda} h^2 = 0.343$, $Y_{\rm He} = 0.24$ and $N_{\nu, \rm eff} = 3.04$.

\section{Hydrogen recombination: overview} \label{sec:overview}

\subsection{The effective three-level atom model} \label{sec:Peebles}

The basic process of primordial hydrogen recombination was already well understood in the late sixties. The seminal papers by Peebles \cite{Peebles} and Zeldovich \emph{et al.} \cite{Zeldovich_et_al} established the following picture. Direct recombinations to the ground state are highly inefficient, as they produce photons that can immediately ionize another hydrogen atom. Electrons and protons can therefore recombine efficiently only to the excited states of hydrogen. This situation is familiar in the study of the interstellar medium: it is referred to as ``case-B'' recombination (see e.g. Ref.~\cite{ISM_book}). Once they have recombined to one of the excited states of hydrogen, electrons ``cascade down'' to the $n=2$ shell, on a much shorter timescale than the overal recombination timescale. Denoting $n_{\rm H}$ the total number density of hydrogen, $x_e = n_e/n_{\rm H}$ the free electron fraction and $x_2 = n_{\textrm{H}(n=2)}/n_{\rm H}$ the fraction of hydrogen in the excited state, the effective rate of recombinations to $n=2$ shell can be written:
\beq
\dot{x}_2\big{|}_{\rm rec} = - \dot{x}_e = n_{\rm H} x_e^2 \alpha_{\rm B}(\Tm) - x_2 \beta_{\rm B}(\Tr), \label{eq:xedot}
\eeq
where $\Tm$ is the matter temperature, locked to the radiation temperature $\Tr$ by Thomson scattering at most times during recombination, $\alpha_{\rm B}$ is the case-B recombination coefficient, and $\beta_{\rm B}$ is the corresponding photoionization rate, which can be obtained from $\alpha_{\rm B}$ by the principle of detailed balance:
\beq
\beta_{\rm B}(\Tr) = \frac{g_e}{4} \rme^{E_2/\Tr} n_{\rm H}\alpha_{\rm B}(\Tm = \Tr), \label{eq:betaB db}
\eeq 
where we have defined
\beq
g_e \equiv \frac{(2 \pi \mu_e \Tr)^{3/2}}{h^3 n_{\rm H}}, \label{eq:ge}
\eeq
where $\mu_e$ is the reduced mass of the electron-proton system.

Once they have reached the $n=2$ shell, electrons can reach the ground state by emitting a Lyman-$\alpha$ photon from the $2p$ state. Due to the very high optical depth of the Lyman-$\alpha$ transition, emitted photons will however almost certainly be reabsorbed by another atom. The way out of this bottleneck is for photons to redshift below the Ly$\alpha$ resonant frequency due to cosmological expansion. The \emph{net} rate of decays to the ground state from the $2p$ state is then just the rate at which photons redshift across the line and escape reabsorption:
\beq
\dot{x}_{1s}\big{|}_{2p} = - \dot{x}_{2p}\big{|}_{1s} = R_{\Ly \alpha} \left( x_{2p} - 3 x_{1s} \rme^{-E_{21}/\Tr}\right), \label{eq:dotx2p simple}
\eeq
where $x_{1s}$ is the fraction of hydrogen in the ground state and the second term accounts for Ly$\alpha$ absorptions and is obtained by detailed balance. The rate of \emph{escape} of Ly$\alpha$ photons is given by:
\beq
 R_{\Ly \alpha} \equiv \frac{8 \pi H}{3 n_{\rm H} x_{1s} \lambda_{\Ly \alpha}^3}.\label{eq:RLya}
\eeq
Eqs.~(\ref{eq:dotx2p simple}--\ref{eq:RLya}) can be derived in the Sobolev approximation, in the limit of large Sobolev optical depth (see for example Ref.~\cite{AGH10}).

The escape rate of Ly$\alpha$ photons is comparable to the rate of the slow two-photon decays from the $2s$ state, $\Lambda_{2s,1s} \approx 8.22$ s$^{-1}$, and the latter process must therefore be accounted for. The net rate of two-photon decays from the $2s$ state is:
\beq
\dot{x}_{1s}\big{|}_{2s} = -\dot{x}_{2s}\big{|}_{1s} = \Lambda_{2s,1s}\left(x_{2s} - x_{1s} \rme^{-E_{21}/\Tr}\right),\label{eq:dotx2s simple}
\eeq
where the second term accounts for two-photon absorptions and can be obtained by a detailed balance argument. Due to the strong thermal radiation bath, the excited states of hydrogen are near Boltzmann equilibrium with each other, $x_{2p} = 3 x_{2s} = (3/4) x_2$. The rate of change of the population of the $n=2$ shell due to decays to the ground state is therefore:
\beq
\dot{x}_2\big{|}_{1s} = \left(\frac34 R_{\Ly \alpha} + \frac14 \Lambda_{2s,1s}\right) \left(4 x_{1s} \rme^{-E_{21}/\Tr} - x_2\right).
\label{eq:dotx2|1s}
\eeq
The last step is to realize that the atomic rates, even for the slow $2s \rightarrow 1s$ decays or the slow escape out of the Ly$\alpha$ resonance, are many orders of magnitude larger than the overall recombination rate, which is of the order of (10 times) the Hubble expansion rate, that is $\sim 10^{-13} -10^{-12}$ s$^{-1}$. The population of the $n=2$ shell can therefore be obtained to high accuracy in the steady-state approximation, i.e. assuming that the rate of recombinations to the $n=2$ shell equals the rate of transitions to the ground state:
\beq
\dot{x}_2 = \dot{x}_2 \big{|}_{\rm rec} + \dot{x}_2\big{|}_{1s} \approx 0.
\eeq 
We can therefore solve for $x_2$ and obtain:
\beq
x_2 = \frac{n_{\rm H} x_e^2 \alpha_{\rm B} +\left(3 R_{\Ly \alpha} +  \Lambda_{2s,1s}\right)  x_{1s} \rme^{-E_{21}/\Tr}} 
      {\beta_{\rm B}+ \frac34 R_{\Ly \alpha} + \frac14 \Lambda_{2s,1s}} \label{eq:Peebles x2}
\eeq
From Eq.~(\ref{eq:xedot}) we then obtain the rate of change of the free electron fraction:
\beq
\dot{x}_e = -C \left(n_{\rm H} x_e^2 \alpha_{\rm B} - 4 x_{1s} \beta_{\rm B} \rme^{- E_{21}/\Tr} \right), \label{eq:Peebles ODE}
\eeq
where the Peebles $C$-factor is given by
\beq
C \equiv \frac{\frac34 R_{\Ly \alpha} + \frac14 \Lambda_{2s,1s}}{\beta_{\rm B} + \frac34 R_{\Ly \alpha} + \frac14 \Lambda_{2s,1s} }. \label{eq:Peebles C}
\eeq
As noted by Peebles, this factor represents the probability that an atom initially the $n=2$ shell reaches the ground state before being photoionized.
Note that we could have obtained the same equation starting from $\dot{x}_e = - \dot{x}_{1s} = -(\dot{x}_{1s}|_{2p} +\dot{x}_{1s}|_{2s})$ (this is because we have set $\dot{x}_2 = 0$).   

At all relevant times during the epoch of hydrogen recombination, $x_2 \ll 1$, and therefore $x_{1s} = 1 - x_e$. If matter and radiation temperatures are set to be equal, Eq.~(\ref{eq:Peebles ODE}) is therefore a simple ordinary differential equation for $x_e$, that can be easily integrated. A simple improvement is to also explicitly follow the matter temperature evolution, which is determined by the Compton evolution equation:
\beq
\dot{T}_{\rm m} = - 2 H \Tm + \frac{8 \sigma_{\rm T} a_{\rm r} \Tr^4 x_e (\Tr - \Tm)}{3 (1 + f_{\rm He} + x_e) m_e c}, \label{eq:Tmdot}
\eeq
where $\sigma_{\rm T}$ is the Thomson cross-section, $a_{\rm r}$ is the radiation constant, $m_e$ is the electron mass and $f_{\rm He}$ is the He:H ratio by number of nuclei. 

The simple yet insightful picture presented here is known as the effective three-level atom model. It provides a good approximation for the recombination problem. It is however not sufficiently accurate for high-precision cosmology. 

\subsection{Hydrogen recombination phenomenology} \label{sec:error impact}

\begin{figure} 
\includegraphics[width = 85mm]{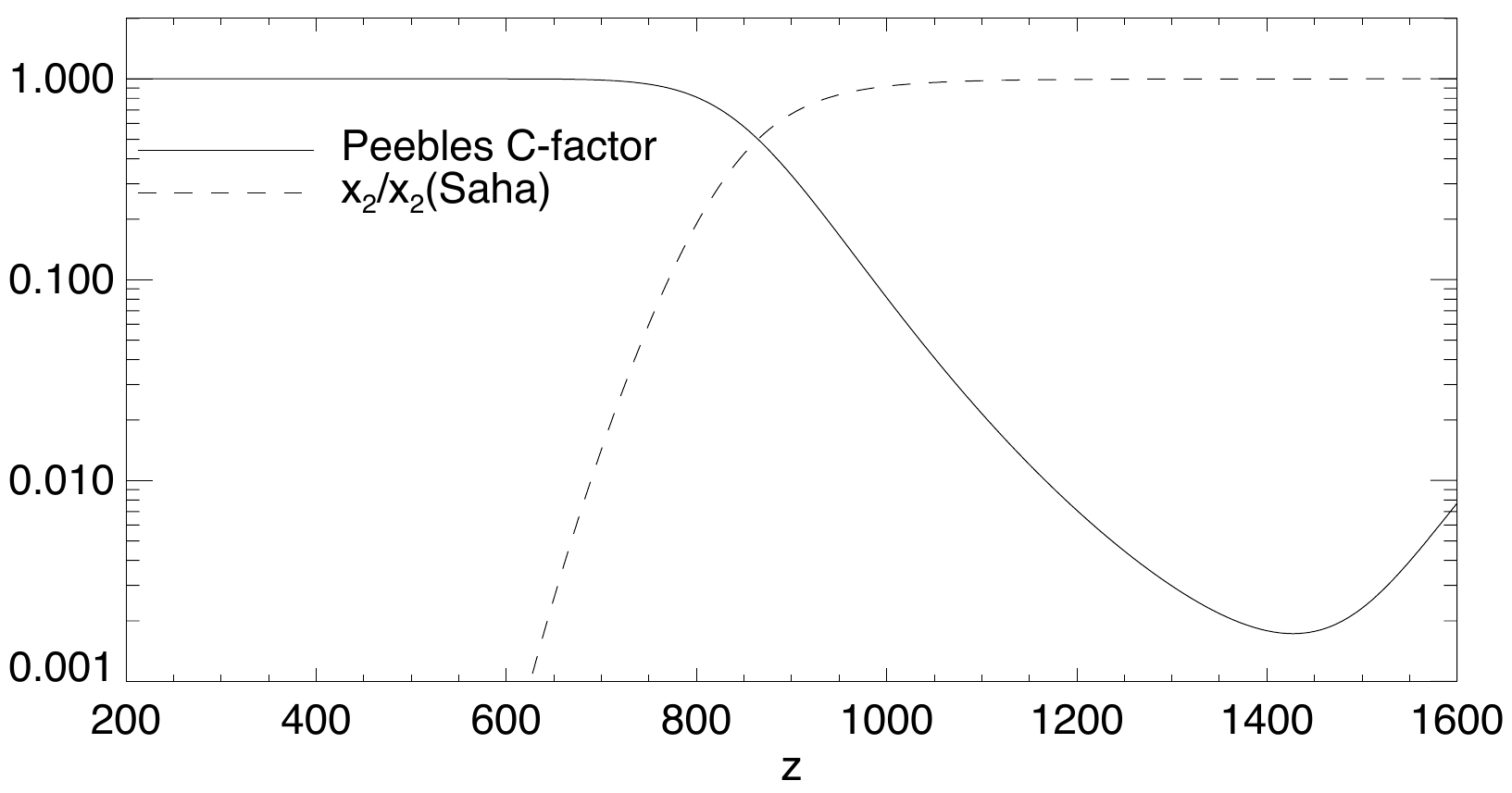}
\caption{Peebles $C$-factor [Eq.~(\ref{eq:Peebles C})] and ratio of the population of the $n=2$ shell to its value in Saha equilibrium with the continuum, as a function of redshift.}
\label{fig:Peebles}
\end{figure}

We show in Fig.~\ref{fig:Peebles} the evolution of the Peebles $C$-factor and the population of the $n=2$ shell relative to its value in Saha equilibrium with the continuum, as a function of redshift, for a standard recombination history. We can see that there are two distinct regimes. 

At early times ($z \gtrsim 1000$), electrons in the $n=2$ shell have a high probability of being photoionized, and the $C$-factor is much smaller than unity, $C \ll 1$. As a consequence, the population of the $n=2$ shell is very close to Saha equilibrium with the continuum, 
\beq
x_2 \approx x_2\big{|}_{\rm Saha} \equiv \frac{4 }{g_e} \rme^{-E_2/\Tm} x_e^2. \label{eq:x2 early}
\eeq
The rate of change of the free electron fraction is then approximately equal to the rate of decays from the $n=2$ shell:
\beq
\dot{x}_e (z \gtrsim 1000) \approx \dot{x}_2\big{|}_{1s}\left(x_2 = x_2\big{|}_{\rm Saha}\right). 
\eeq
During that period, the recombination rate is therefore virtually independent of the exact value of the recombination coefficient, but is strongly dependent on the small net decay rate from the $n=2$ shell to the ground state. This is usually referred to as the ``$n=2$ bottleneck'' and has motivated abundant work on radiative transfer in the vicinity of the Lyman transitions \cite{CS09a, CS07, CS_2photon, Hirata_2photon, CS09b, Hirata_Forbes, CS09c, AGH10, CS06, Kholupenko06, Kholupenko_Deuterium}. The result from this series of papers is that to the level of accuracy required by \emph{Planck}, Lyman transitions up to Ly$\gamma$ must be included, properly accounting for feedback between them. In addition, the radiation field must be solved for with a radiative transfer calculation, accounting for two-photon transitions and frequency diffusion in the Lyman-$\alpha$ line. We consider all these effects in Sections~\ref{sec:two-photon formal} and \ref{sec:two-photon numerical}.

At late times ($z \lesssim 700$), $C \approx 1$, and the $n=2$ shell is no longer in Saha equilibrium with the continuum (note that it is \emph{not} in Boltzmann equilibrium with the ground state either, as the rate of recombinations to the $n=2$ shell dominates over the net rate of two-photon or Ly-$\alpha$ absorptions from the ground state). The free electron fraction is many orders of magnitude above the value it would have in Saha equilibrium because of the slow recombination rate. In that case, the second term in Eq.~(\ref{eq:Peebles ODE}) is negligible and the evolution of the free electron fraction becomes:
\beq
\dot{x}_e (z \lesssim 700) \approx - n_{\rm H} x_e^2 \alpha_{\rm B}.
\eeq
As we can see, the evolution of the free electron fraction is then virtually independent of the rate of decays to the ground state from the $n=2$ shell, but is highly sensitive to the exact value of the effective recombination coefficient. Moreover, the assumption that the excited states are in Boltzmann equilibrium with each other brakes down at late times because of the decrease in the radiation temperature. An accurate recombination rate at late times can only be obtained in a full multilevel atom calculation, that accounts for (possibly stimulated) bound-bound and bound-free transitions between all -- at least, a large number of -- the excited states of hydrogen \cite{Recfast_long, CRMS07, Grin_Hirata, Chluba_Vasil}. We will review the multi-level atom calculations in Sec.~\ref{sec:MLA}.

Of course, at intermediate redshits $700 \lesssim z \lesssim 1000$ both the exact recombination rate and the rate of decays to the ground state are important and should be carefully accounted for.

\section{The multi-level atom} \label{sec:MLA}

In this section we first present the ``standard'' multi-level atom (MLA) method \cite{Recfast_long}, then review the effective MLA (hereafter EMLA) method of solution that we presented in Paper I and that allows for a fast computation of recombination histories. 

\subsection{The standard multi-level atom method}

The effective three-level atom equations may be easily generalized to account for an arbitrarily large number of excited states of hydrogen (in practice, one must of course impose a cutoff). We denote $x_{nl}$ the fractional abundance of hydrogen atoms in the excited state with principal quantum number $n$ and angular momentum quantum number $l$. The generalization of Eq.~(\ref{eq:xedot}) is then, for $n \geq 2$:
\beq
\dot{x}_{nl}\big{|}_{\rm rec} = n_{\rm H} x_e^2 \alpha_{nl}(\Tm, \Tr) - x_{nl} \beta_{nl}(\Tr),  
\eeq
where $\alpha_{nl}(\Tm, \Tr)$ is the recombination coefficient to the excited state $nl$, including stimulated recombinations, and $\beta_{nl}(\Tr)$ is rate of photoionizations from $nl$ by blackbody photons. 

The effective three-level atom model does not account for bound-bound transitions between excited states (except for instantaneous spontaneous decays that ultimately lead to the $n=2$ shell). Transitions between the $nl$ and $n'l'$ states (with $n, n' \geq 2$) change their populations at the rate:
\barr
\dot{x}_{nl}\big{|}_{n'l'} &=& - \dot{x}_{n'l'}\big{|}_{nl} \nonumber \\
&=& x_{n'l'} R_{n'l' , nl}(\Tr) - x_{nl} R_{nl , n'l'}(\Tr),
\earr
where the bound-bound transition rate from $nl$ to $n'l'$, $R_{nl , n'l'}(\Tr)$, is the rate of absorptions of blackbody photons resonant with the transition if $n < n'$ and the rate of spontaneous and stimulated decays if $n > n'$. We give explicit expressions of the bound-bound and bound free rates and explain how we compute them in Paper I.

Finally, the rate of decays to the ground state from the $2s$ state are given by Eq.~(\ref{eq:dotx2s simple}) (we will see how to make this rate more accurate in Sec.~\ref{sec:twoG}). In the Sobolev approximation, the net decay rate from the $np$ states to the ground state is given by a generalization of Eq.~(\ref{eq:dotx2p simple}), accounting for feedback between optically thick Lyman lines:
\beq
\dot{x}_{1s}\big{|}_{np} = -\dot{x}_{np}\big{|}_{1s} = R_{\Ly n}\left[x_{np} - 3 x_{1s} f_{np}^+\right], \label{eq:dotxnp feedback}
\eeq
where $R_{\Ly n} \equiv (\lambda_{\Ly \alpha}/\lambda_{\Ly n})^3 R_{\Ly \alpha}$ is the rate at which photons redshift out of the Lyman-$n$ line (with the convention that Ly-$2$ is Ly$\alpha$), and $f_{np}^+$ is the photon occupation number incoming on the blue side of the Ly-$n$ transition. If no radiative processes affect the radiation field between neighboring Lyman lines, then
\beq
f_{np}^+(z) = f_{n+1,p}^-(z'), 
\eeq
where the earlier redshift $z'$ is given by
\beq
z' = \frac{\lambda_{\Ly n}}{\lambda_{\Ly (n+1)}}(1 + z) - 1.
\eeq 
In the optically thick limit which is valid here, the photon occupation number redward of the Ly-$n$ line is given by $f_{np}^- = x_{np}/(3 x_{1s})$.

For excited states $nl$ other than $2s$ and $np$, not radiatively connected to the ground state, we have $\dot{x}_{nl}\big{|}_{1s} = 0$.

The recombination history can then be computed by evolving simultaneously the system of differential equations:
\barr
\dot{x}_{nl} &=& \dot{x}_{nl}\big{|}_{\rm rec} + \sum_{n'\geq 2, l'} \dot{x}_{nl}\big{|}_{n'l'} + \dot{x}_{nl}\big{|}_{1s} \label{eq:dot xnl}\\
\dot{x}_e &=& - \dot{x}_{1s} = \dot{x}_{2s}\big{|}_{1s} + \sum_{n \geq 2} \dot{x}_{np}\big{|}_{1s}, \label{eq:xedot MLA}
\earr
where in the last equation we used $x_e = 1 - x_{1s}$, valid as the fractional abundance of hydrogen in the excited states is always much less than unity. 
The atomic transition rates are many orders of magnitude larger than the overall recombination rate (of the order of the Hubble rate). A highly accurate approximation therefore consists in first solving for the populations of the excited states in the steady-state approximation. This first step amounts to solving a large system of linear algebraic equations. The populations $x_{2s}, x_{np}$ can then be used in Eq.~(\ref{eq:xedot MLA}) to evolve the free electron fraction.

This generalization of the three-level atom model is relatively straightforward conceptually. Its practical implementation is, however, very time-consuming. It requires solving a very large system of algebraic equations at each time step (or evolving the same number of stiff differential equations). If one accounts for excited states up to principal quantum number $n_{\max}$, then the number of equations is $N = n_{\max}(n_{\max} + 1)/2$, which, for $n_{\max} \gtrsim 100$, exceeds several thousands. Furthermore, the computational cost of an exact linear system solution scales as ${\cal O}(N^3)$, although this can be significantly sped up by using the sparseness of the system due to selection rules \cite{Grin_Hirata} and iterative solution techniques \cite{CVD10}.  In the following section, we review the much more efficient yet exactly equivalent effective MLA method.

\subsection{The effective multi-level atom method}
\subsubsection{General description}

The effective multilevel atom method, described in Paper I, relies on three aspects of the primordial recombination problem. First, the timescales for transitions out of the excited states are much shorter than the overall recombination timescale -- this property is used when solving for the populations of the excited states in the steady-state approximation in the standard MLA method. This allows us to factor all the \emph{nearly instantaneous} transitions involving the ``interior'' excited states (which are not radiatively connected to the ground state) into effective transitions into and out of the smaller set of ``interface'' states which are radiatively connected to the ground state. Secondly, all bound-bound and bound-free transitions for which the lower state is an excited state are \emph{optically thin}, and therefore do not distort the ambient blackbody radiation field in the vicinity of the corresponding frequencies. All the transition rates between ``interior'' states therefore only depend on the radiation temperature $\Tr$ (as well as atomic physics constants). This translates into a simple dependence for the effective rates, which are functions of the matter (through recombinations of thermal electrons and protons) and radiation temperatures only. If collisional transitions are included, they will depend additionally on the free electron (of equivalently the free proton) abundance. Finally, the set of ``interface'' states that need to be considered is small. In principle, $2s$ and all the $p$ states need to be considered as ``interface'' states; however, in practice only the lowest order Lyman transitions significantly affect the recombination rate, explicitly only Ly$\alpha$, Ly$\beta$ and Ly$\gamma$ \cite{EMLA, AGH10}. In addition, when considering two-photon transitions from higher levels, one should in principle add the $ns$ and $nd$ states as ``interface'' states. However, only two-photon transitions from $2s$, $3s$ and $3d$, are important (and $4s$, $4d$ at the level of a few $10^{-4}$) \cite{Hirata_2photon, CS09b}. This means that one needs to pretabulate only \emph{a few functions} of temperature that fully account for the multilevel structure of hydrogen. These tabulated effective rates can then be interpolated when computing a recombination history with an effective few-level atom model. As we will show in Sec.~\ref{sec:E4LA}, we can in fact further simplify the problem to an effective four-level atom model with virtually no loss of accuracy. 

Following Paper I, we denote $\mathcal{A}_i(T_{\rm m}, T_{\rm r})$ the effective recombination coefficient to the ``interface'' state $i$, $\mathcal{B}_i(T_{\rm r})$ the effective photoionization rate from this state, and $\mathcal{R}_{i , j}(\Tr)$ the transfer rate from the interface state $i$ to the interface state $j$ (the dependences are valid in the purely radiative case; when collisions are included, all effective rates depend on $\Tm, \Tr ,n_e$). They are obtained as follows: for the effective recombination coefficients,
\beq
\mathcal{A}_i = \alpha_i + \sum_K \alpha_K P_K^i; \label{eq:Ai}
\eeq
for the effective ionization coefficient,
\beq
\mathcal{B}_i = \beta_i + \sum_K R_{i , K} P_K^e;  \label{eq:Bi}
\eeq
and for the effective inter-state transition rates,
\beq
\mathcal{R}_{i , j} = R_{i , j} + \sum_K R_{i , K} P_K^j. \label{eq:Rij}
\eeq
Here $K$ is a general index for ``interior'' states, $P_K^i$ is the probability that an electron initially in the ``interior'' state $K$ ultimately reaches the ``interface'' state $i$, and $P_K^e$ is the probability that an atom initially in the state $K$ ultimately gets photoionized. These probabilities are the solutions of the linear systems \cite{EMLA}:
\beq
P_K^i = \sum_L \frac{R_{K , L}}{\Gamma_K} P_L^i + \frac{R_{K , i}}{\Gamma_K} \label{eq:PKi}
\eeq
and
\beq
P_K^e = \sum_L \frac{R_{K , L}}{\Gamma_K} P_L^e + \frac{\beta_K}{\Gamma_K}, \label{eq:PKe}
\eeq
where 
\beq
\Gamma_K \equiv \sum_L R_{K , L} + \sum_i R_{K , i} + \beta_K \label{eq:GammaK}
\eeq
is the inverse lifetime of the state $K$.

The \emph{net} decay rate from the interface state $i$ to the ground state can always be expressed as a linear function of $x_i$:
\beq
\dot{x}_{1s}\big{|}_{i} = - \dot{x}_i\big{|}_{1s} =  x_i \tilde{R}_{i,1s} - x_{1s} \tilde{R}_{1s , i},
\eeq
where we emphasize with this notation that in general the net decay rates may depend in a complicated way on the current and past values of the free electron fraction, as well as on cosmological parameters -- see for example Eqs.~(\ref{eq:dotxnp feedback}) and (\ref{eq:RLya}) for the net $np \rightarrow 1s$ decay rate. This contrasts with the bound-bound rates between excited states, which only depend on atomic constants and the radiation temperature.

Once the effective bound-bound and bound-free rates for the interface states are tabulated, the recombination history can be computed by evolving the small set of ordinary differential equations:
\barr
\dot{x}_i &=& x_e^2 n_{\rm H} \mathcal{A}_i + \sum_{j \neq i} x_j \mathcal{R}_{j , i} + x_{1s} \tilde{R}_{1s , i} \nonumber\\
 &&- x_i\left( \mathcal{B}_i + \sum_{j\neq i} \mathcal{R}_{i , j} + \tilde{R}_{i , 1s}\right)\label{eq:xidot} 
\earr
and
\barr
\dot{x}_e &=& - \sum_i \left( n_{\rm H} x_e^2 \mathcal{A}_i - x_i \mathcal{B}_i\right)\label{eq:xedot EMLA}\\
&=& - \dot{x}_{1s} = \sum_i\left(x_{1s}\tilde{R}_{1s,i} - x_i \tilde{R}_{i,1s}\right),\label{eq:x1sdot EMLA}
\earr
where Eqs.~(\ref{eq:xedot EMLA}) and (\ref{eq:x1sdot EMLA}) are equivalent as the fractional abundance of excited hydrogen is very small. In practice, the population of the effective states can once again be solved in the steady-state approximation. Eq.~(\ref{eq:xidot}) becomes a system of a few algebraic linear equations, and in that case Eqs.~(\ref{eq:xedot EMLA}) and (\ref{eq:x1sdot EMLA}) are mathematically equivalent since we set $\dot{x}_i = 0$ [this can be seen by summing Eq.~(\ref{eq:xidot}) over $i$].

Equations (\ref{eq:xidot}) and (\ref{eq:xedot EMLA}), along with the definitions for the effective coefficients Eqs.~(\ref{eq:Ai}--\ref{eq:GammaK}), are strictly equivalent to the standard MLA equations presented in the previous section, as was derived in Paper I [Eq.~(\ref{eq:xedot EMLA}) was not derived in that paper and we give a proof in Appendix \ref{app:xedot}]. The advantage of the new method is that the system of equations that need to be solved at each time step is much smaller, as only the low-lying $s, p, d$ states need to be followed. In the following section we show that we can actually further reduce the problem to an effective four-level atom model.

\subsubsection{Further simplification: the effective four-level atom}\label{sec:E4LA}

If we wish to follow $n_*$ ``interface'' states, then the system of equations (\ref{eq:xidot}), in the steady-state approximation, is a $n_* \times n_*$ system. Moreover, one needs to interpolate $n_*$ functions of 2 variables (the effective recombination coefficients -- the effective photoionization rates are obtained by detailed balance), and $n_*(n_* - 1)/2$ functions of 1 variable (half of the effective bound-bound rates, the other half being obtained by detailed balance). Here we show how this system can be further reduced to a $2 \times 2$ system involving only $2s$ and $2p$, requiring only 2 functions of 2 variables and $2n_* - 3$ functions of one variable, with virtually no loss of accuracy. 

For now on we use the general index $K$ for all states with principal quantum number $n \geq 3$. Even if some of the states with $n \geq 3$ are radiatively connected to the ground state, one can still formally define the effective transition rates Eqs.~(\ref{eq:Ai}), (\ref{eq:Bi}) and (\ref{eq:Rij}) for the $2s$ and $2p$ states because of the near-instantaneity of transitions out of the excited states. However, these coefficients do not have a simple temperature dependence anymore, and are therefore not well suited for fast interpolation. Indeed, the probabilities $P_K^i$ and $P_K^e$, where $i = 2s, 2p$, are still defined by Eqs.~(\ref{eq:PKi}) and (\ref{eq:PKe}), but the inverse lifetime of the state $K$, Eq.~(\ref{eq:GammaK}), should now account for the net downward transition rate to the ground state, and one should make the replacement:
\beq
\Gamma_K \rightarrow \tilde{\Gamma}_K \equiv \Gamma_K(\Tr) + \tilde{R}_{K,1s}, \label{eq:tilde GammaK}
\eeq
where we define $\Gamma_K(\Tr)$ is the inverse lifetime of the $K$-th interface state when transitions to the ground state are not included.
In addition, we need to define additional effective transition rates with the ground state. For $i = 2s, 2p$, we define:
\beq
\tilde{\mathcal{R}}_{i,1s} \equiv \tilde{R}_{i,1s} + \sum_K R_{i,K} \tilde{P}_K^{1s} \label{eq:Ri1s}
\eeq
and
\beq
\tilde{\mathcal{R}}_{1s,i} \equiv \tilde{R}_{1s,i} + \sum_K \tilde{R}_{1s,K} P_K^{i}, \label{eq:R1si}
\eeq
where the probabilities $\tilde{P}_K^{1s}$ must satisfy the self-consistency relations:
\beq
\tilde{P}_K^{1s} = \sum_L \frac{R_{K,L}}{\tilde{\Gamma}_K} \tilde{P}_{L}^{1s} + \frac{\tilde{R}_{K, 1s}}{\tilde{\Gamma}_K}. \label{eq:tildePK1s}
\eeq
The standard MLA equations, in the steady-state approximation for the excited states with $n \geq 3$, can then be shown to be exactly equivalent to the following set of equations: for the net rate of production of $2s$,
\barr
\dot{x}_{2s} &=& x_e^2 n_{\rm H} \mathcal{A}_{2s} + x_{2p} \mathcal{R}_{2p , 2s} + x_{1s} \tilde{\mathcal{R}}_{1s , 2s} \nonumber\\
 &&- x_{2s}\left( \mathcal{B}_{2s} + \mathcal{R}_{2s , 2p} + \tilde{\mathcal{R}}_{2s , 1s}\right);\label{eq:x2sdot} 
\earr
for the net rate of production of $2p$,
\barr
\dot{x}_{2p} &=& x_e^2 n_{\rm H} \mathcal{A}_{2p} + x_{2s} \mathcal{R}_{2s , 2p} + x_{1s} \tilde{\mathcal{R}}_{1s , 2p} \nonumber\\
 &&- x_{2p}\left( \mathcal{B}_{2p} + \mathcal{R}_{2p , 2s} + \tilde{\mathcal{R}}_{2p , 1s}\right);\label{eq:x2pdot} 
\earr
and for either the net recombination rate or the net rate of generation of H(1s),
\barr
\dot{x}_e &=& \sum_{i=2s,2p} \left[x_{i}\mathcal{B}_{i} - x_e^2 \nh \mathcal{A}_{i}\right] \label{eq:xedot EFLA}\\
&=& \sum_{i=2s,2p} \left[x_{1s}\tilde{\mathcal{R}}_{1s,i} - x_{i}\tilde{\mathcal{R}}_{i,1s}\right].\label{eq:x1sdot EFLA}
\earr
The proof of equivalence is a simple generalization of that of Paper I and we do not reproduce it here. Note that we use the same notation for the effective rates independently of the number of ``interface'' states considered, but they obviously have a different meaning that should be clear from the context.

The coefficients in the above system are in principle not simple functions of temperature anymore if one wishes to account for the transitions to the ground state from excited states with $n \geq 3$. We can nevertheless simplify their expressions with some minimal approximations. We start by noticing that for excited states $nl$ with $n \geq 3$, the rate of spontaneous decays to the $n', l\pm 1$ states with $1 < n' < n$ is much larger than the net decay rate to the ground state. For the $3p$ state for example, we find that in the Sobolev approximation for Ly$\beta$ decays, $\tilde{R}_{3p,1s}/A_{3p,2s} < 8 \times 10^{-6}$ for $200 < z < 1600$, where $A_{3p,2s}$ is the Einstein A-coefficient of the $3p \rightarrow 2s$ transition. Therefore, to an excellent accuracy (with relative errors of order $\tilde{R}_{K,1s}/\Gamma_K$), one can neglect $\tilde{R}_{K,1s}$ in Eq.~(\ref{eq:tilde GammaK}), and simply use $\Gamma_K(\Tr)$ instead of $\tilde{\Gamma}_K$ wherever the latter appears. With this approximation, the rate coefficients $\mathcal{A}_{2s/2p}$, $\mathcal{B}_{2s/2p}$ and $\mathcal{R}_{2s, 2p}$ are simply the usual effective rates computed in the case that only $2s$ and $2p$ are considered as interface states, and depend only on matter and radiation temperatures. We explain in Appendix \ref{app:extrapolation} how we obtain effective rates extrapolated to $n_{\max} = \infty$.

We show in Appendix \ref{app:proof Ri1s} that using $\tilde{\Gamma}_K \approx \Gamma_K(\Tr)$ in Eq.~(\ref{eq:tildePK1s}), we can rewrite Eq.~(\ref{eq:Ri1s}) as:
\beq
\tilde{\mathcal{R}}_{i,1s} = \tilde{R}_{i,1s} + \sum_K \tilde{R}_{K, 1s} \frac{g_K}{g_i} \rme^{-E_{K2/\Tr}} P_K^i(\Tr),\label{eq:Ri1s-new}
\eeq
 where $g_K, g_i$ are the statistical weights of the states $K, i$ and $E_{K2} \equiv E_K - E_2$ is the energy difference between the state $K$ and the $n = 2$ shell.

In addition, the populations of the ``weak'' interface states $X_K$ are sometimes required -- for example the photon occupation number depends on the populations of the $s$ and $d$ states, see Section \ref{sec:rad trans sol}. We show in Appendix \ref{app:XK} that the following relation is verified for $\Tm = \Tr$:
\barr
X_K\big{|}_{\Tm = \Tr} &=& \frac{g_K}{g_e} \rme^{-E_K/\Tr} P_K^e(\Tr)  x_e^2\nonumber\\
 &+& \sum_{i = 2s, 2p} \frac{g_K}{g_i} \rme^{-E_{K2}/\Tr} P_K^i(\Tr)  x_i. \label{eq:XK-effective}
\earr
In fact, $\Tm < \Tr$ and the coefficient of $x_e^2$ in the above equation should be slightly higher. In practice though, $|\Tm/ \Tr - 1| < 1\%$ for $z \gtrsim 500$, and for lower redshifts $P_K^e \ll 1$ and transitions to the ground state are unimportant anyway, so the above equation is very accurate.  

We therefore only need to tabulate the additional $2(n_* - 2)$ functions $P_K^{2s} (\Tr)$ and $P_{K}^{2p}(\Tr)$ to account for $n_*-2$ ``weak'' interface states in addition to $2s$ and $2p$ (note that $P_K^e = 1 - P_K^{2s} - P_K^{2p}$). 

In practice, we can further reduce the computational load by simply using $P_{ns}^{2s} = P_{nd}^{2s} = P_{np}^{2p} = 0, ~P_{ns}^{2p} = P_{nd}^{2p} = P_{np}^{2s} = 1$. This amounts to assuming that for $n\geq 3$, $ns$ and $nd$ states are in Boltzmann equilibrium with $2p$, whereas $np$ states are in Boltzmann equilibrium with $2s$ -- see Eq.~(\ref{eq:XK-effective}), and rewriting transitions from $n \geq 3$ states as transitions from the $n=2$ state to which they are tightly coupled. This is extremely accurate at all times when two-photon transitions significantly affect recombination. The validity of this statement is somewhat weaker for the $n=4$ states, which are somewhat in between equilibrium with the $2s$ state and the $2p$ state (because allowed transitions connect them to both states). However, as decays from $4p$, $4s$ and $4d$ only marginally affect recombination anyway (at the level of a few $10^{-4}$), and $2s$ and $2p$ are very close to equilibrium at the relevant times, this approximation is still very accurate. We explicitly checked that using the approximate values for the $P_K^i$ instead of their exact values in Eqs.~(\ref{eq:R1si}), (\ref{eq:Ri1s-new}) and (\ref{eq:XK-effective}) leads to maximum errors on the recombination history $|\Delta x_e|/x_e < 3 \times 10^{-5}$. 

\begin{figure} 
\includegraphics[width = 85mm]{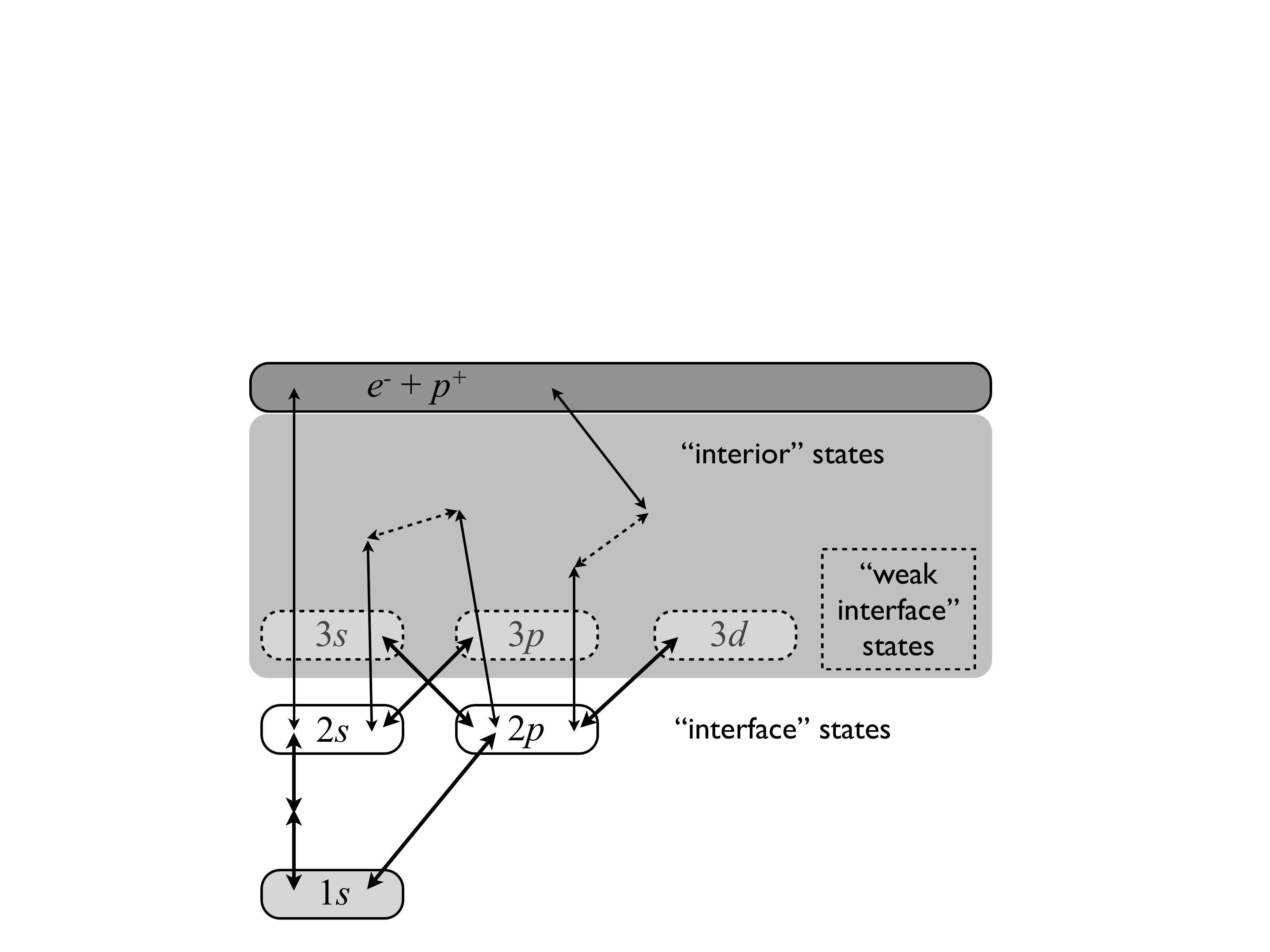}
\caption{Schematic representation of the hydrogen atom, with the nomenclature used in this paper. Slow transitions from the ``weak interface'' states to the ground state are counted as transitions from the $n=2$ state with which they are in equilibrium.}
\label{fig:scheme}
\end{figure}

We illustrate the formulation adopted in this paper schematically in Fig.~\ref{fig:scheme}.

\begin{figure} 
\includegraphics[width = 85mm]{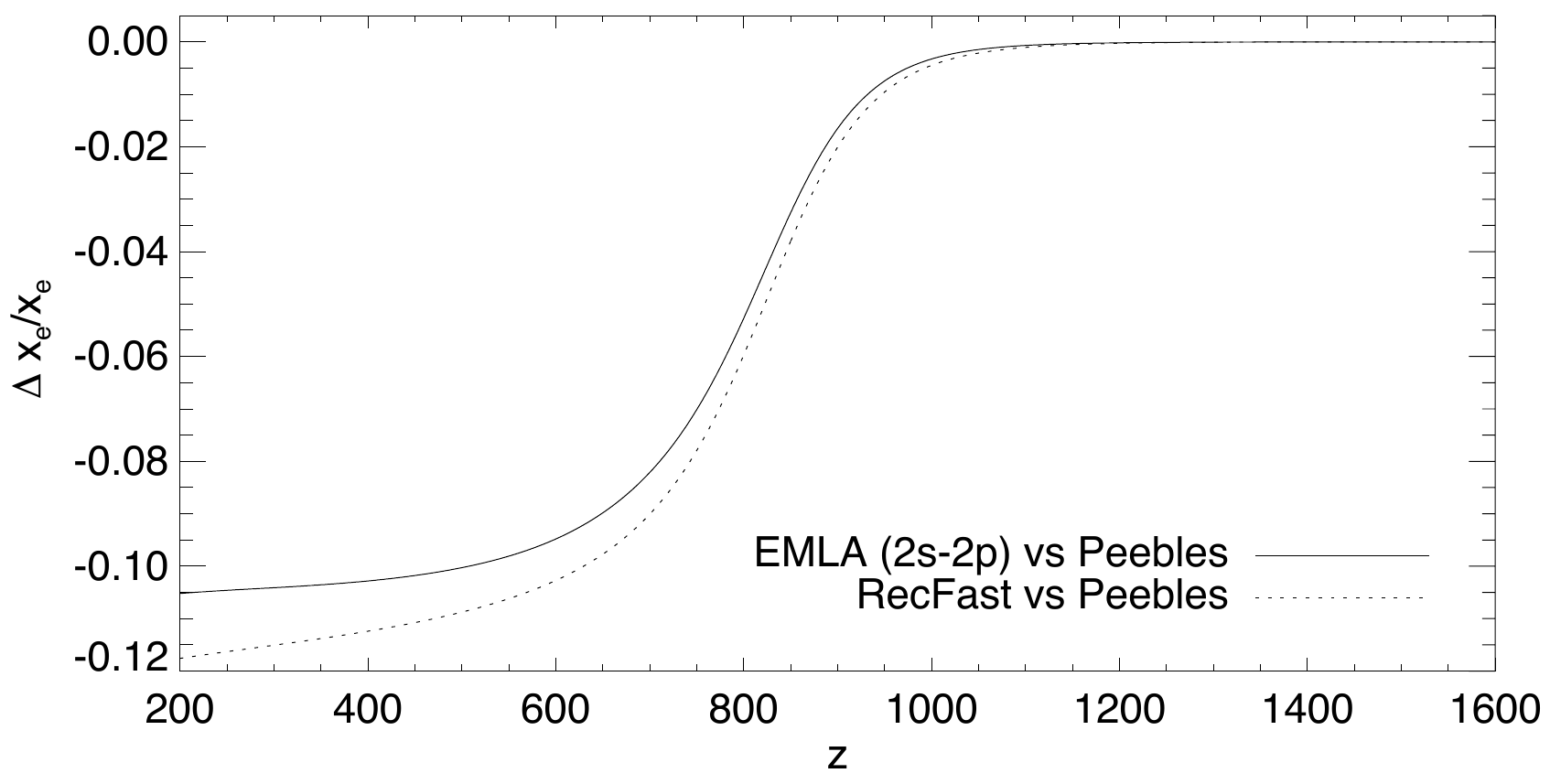}
\caption{Fractional changes in the ionization history relative to the effective three-level atom model. The ``RecFast'' model is an effective three-level atom with the case-B recombination coefficient multiplied by a fudge factor F = 1.14. The same prescription for the evolution of the matter temperature is used in all cases, see Sec.\ref{sec:results}.}
\label{fig:peebles_compare}
\end{figure}

The system of equations (\ref{eq:x2sdot}), (\ref{eq:x2pdot}) and (\ref{eq:xedot EFLA}) is just the extension of Peebles effective three-level atom to an effective four-level atom, properly accounting for the non-zero radiation field, the nearly instantaneous multiple transitions between excited states, the fact that $2s$ and $2p$ are out of Boltzmann equilibrium, and possibly additional radiative transfer effects and decays from higher shells through the appropriate coefficients $\tilde{\mathcal{R}}$. Making the usual steady-state assumption for the excited states, we can first solve for $x_{2s}$ and $x_{2p}$ and then evolve $x_e$.  When using the simple $2p\leftrightarrow 1s$ and $2s \leftrightarrow 1s$ transition rates of Sec. \ref{sec:Peebles}, the system is simple enough that we may write the function $\dot{x}_e$ explicitly as an illustration:
\barr
\dot{x}_e = &-& C_{2s}\left(n_{\rm H} x_e^2 \mathcal{A}_{2s} - ~~ x_{1s} \mathcal{B}_{2s}\rme^{-E_{21}/\Tr}\right) \nonumber\\
               &-& C_{2p}\left(n_{\rm H} x_e^2 \mathcal{A}_{2p} -3 x_{1s} \mathcal{B}_{2p}\rme^{-E_{21}/\Tr}\right) , \label{eq:dotxe-explicit-EFLA}
\earr
where the $C$-factors are given by
\beq
C_{2s} \equiv \frac{\Lambda_{2s,1s}+ \mathcal{R}_{2s\rightarrow 2p}\frac{R_{\rm Ly \alpha}}{\Gamma_{2p}}}{\Gamma_{2s} - \mathcal{R}_{2s\rightarrow 2p} \frac{\mathcal{R}_{2p\rightarrow 2s}}{\Gamma_{2p}}},\\
\eeq
and
\beq
C_{2p} \equiv \frac{R_{\rm Ly \alpha}+ \mathcal{R}_{2p\rightarrow 2s}\frac{\Lambda_{2s,1s}}{\Gamma_{2s}}}{\Gamma_{2p} - \mathcal{R}_{2p\rightarrow 2s} \frac{\mathcal{R}_{2s\rightarrow 2p}}{\Gamma_{2s}}},
\eeq
and where we have used the effective inverse lifetimes:
\barr
\Gamma_{2s} &\equiv& \mathcal{B}_{2s}+ \mathcal{R}_{2s,2p} + \Lambda_{2s,1s} {\rm ~~and}\nonumber \\
\Gamma_{2p} &\equiv& \mathcal{B}_{2p}+ \mathcal{R}_{2p,2s} + R_{\rm Ly \alpha}.
\earr
In Fig.~\ref{fig:peebles_compare} we show the changes to the recombination history resulting from an accurate effective multi-level computation, as compared to the effective three-level atom computation, using in both cases the simple decay rates to the ground state described in Sec~\ref{sec:Peebles}. For comparison, we also show the resulting changes when using an effective three-level atom model with a fudge factor $F = 1.14$ as in the code \textsc{RecFast} \cite{Recfast_short}. We checked that it is not possible to reproduce the correct effective MLA computation with a constant fudge factor in an effective three-level atom. We find that the best fitting fudge factor would be $F = 1.126$, with relative errors reaching 0.2 \%. In any case, the effective MLA computation is so simple and computationally efficient that the need for non-physical fudge factors does not arise.

\begin{figure} 
\includegraphics[width = 85mm]{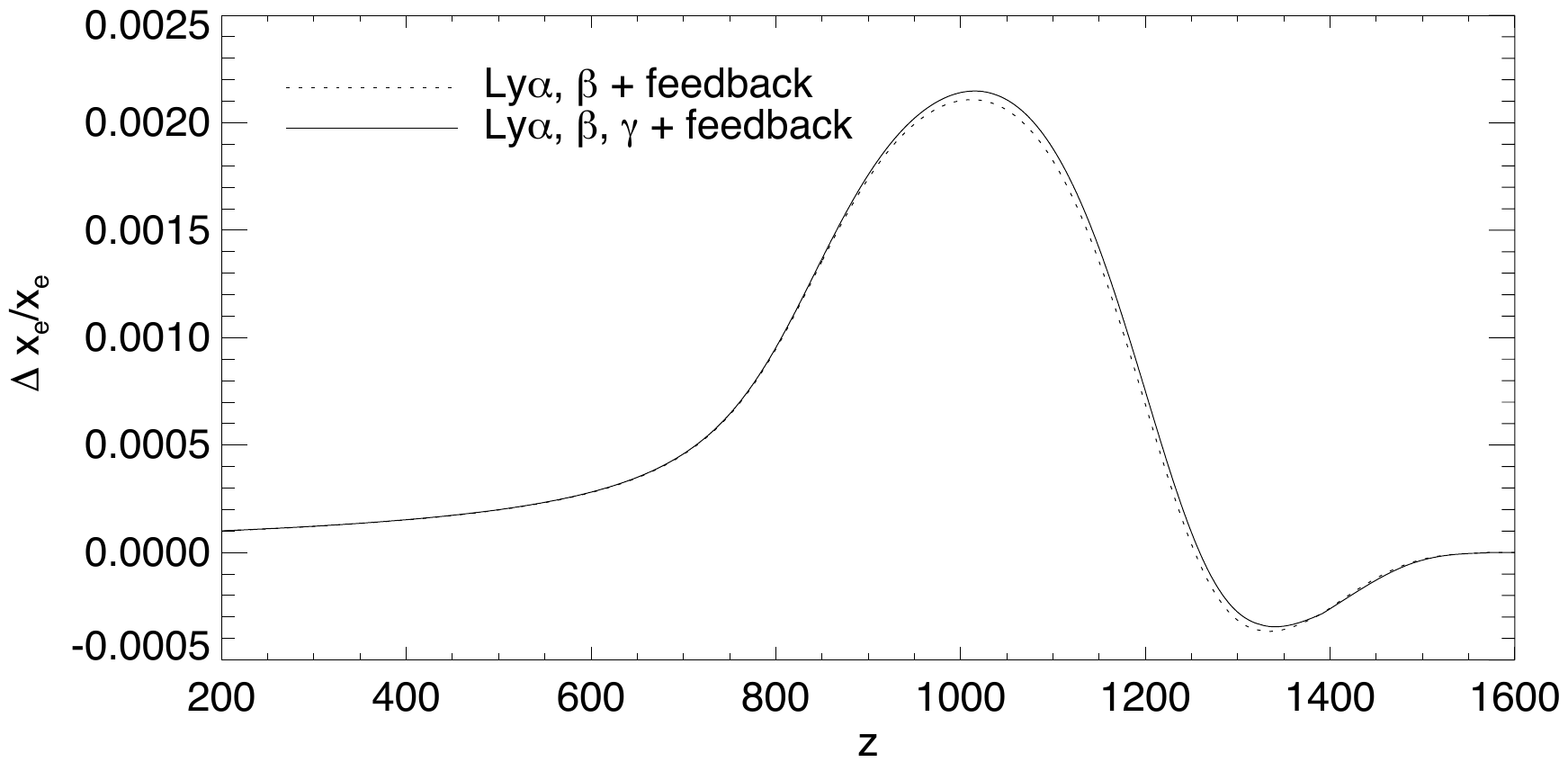}
\caption{Fractional changes in the ionization history when including higher-order Lyman transitions and feedback between them, compared to the effective multi-level atom model with $2s$ and $2p$ only.}
\label{fig:feedback}
\end{figure}

In Fig.~\ref{fig:feedback} we show the effect of adding higher order Lyman transitions and feedback between them. It is sufficient to include Lyman transitions up to Ly$\gamma$, as neglecting higher transitions leads to relative changes of $10^{-5}$ only \cite{AGH10}. This initially speeds up recombination by adding more decay paths to the ground state, then slows it down due to delayed reabsorptions of Ly$\beta$ photons in the Ly$\alpha$ line. Our results are similar to those of Ref.~\cite{Kholupenko_Deuterium}.

For now on, our ``base'' model will be the effective multi-level atom model with Lyman-$\alpha$, $\beta$ and $\gamma$ transitions and feedback between them (assuming a blackbody radiation field incoming on Ly$\gamma$). In the next sections we graft two-photon processes onto this base model. 

\section{Two-photon processes: formal description} \label{sec:two-photon formal}

\subsection{Overview}\label{sec:twoG}

It is well known since the first works on primordial recombination that $2s \rightarrow 1s$ two-photon decays significantly contribute to the recombination dynamics \cite{Peebles, Zeldovich_et_al}. Even with a relatively low decay rate, the forbidden $2s \rightarrow 1s$ decays are indeed comparable in efficiency to the highly self-absorbed Lyman-$\alpha$ transition; in fact, more than half of hydrogen atoms have formed through the $2s\rightarrow 1s$ channel \cite{CS06b}. This process was traditionally accounted for with the total $2s \rightarrow 1s$ decay rate in vacuum, $\Lambda_{2s,1s} \approx 8.22$ s$^{-1}$, with a two-photon absorption rate obtained by detailed balance considerations. For the level of accuracy required for future CMB experiments, one needs to account for stimulated two-photon decays \cite{CS06} and non-thermal absorptions \cite{Kholupenko06, Hirata_2photon}. 

Recently, it was suggested that two-photon decays from higher lying $ns$ and $nd$ states may also lead to percent level corrections to the recombination history \cite{Dubrovich_Grachev05}. Inclusion of such decays presents an additional conceptual difficulty which was not present for the $2s \rightarrow 1s$ decays: the problem of double-counting. Indeed, there is no fundamental difference between a sequence of two allowed one-photon transitions $nl \rightarrow n'p$, $n'p \rightarrow 1s$, with $1 < n' < n$, and a two photon decay from the $nl$ state near resonance (i.e. where the energy of the two photons are near $E_{nn'}$ and $E_{n'1}$ respectively). Approximate solutions were presented in Refs.~\cite{Dubrovich_Grachev05, Wong_Scott07, CS_2photon} (for a review, see Ref.~\cite{Hirata_2photon}). The double-counting problem as well as the reabsorption problem were resolved with a numerical approach, solving the radiative transfer equations for the photon field in Ref.~\cite{Hirata_2photon}, which also provided analytic approximations to check the validity of the numerical result. In this work, we will use the same numerical method as in Ref.~\cite{Hirata_2photon}, which we then extend to account for frequency diffusion near the Ly$\alpha$ line. In this section, we review the formalism presented in Ref.~\cite{Hirata_2photon} and how to solve the double counting problem. In Sec.~\ref{sec:two-photon numerical} we will describe our numerical method for solving simultaneously the radiative transfer equation and the evolution of the atomic level populations. 

\subsection{Two-photon decays and Raman scattering}
We start by defining the coefficient:
\beq
\frac{d \Lambda_{nl}}{d \nu} \equiv \frac{\alpha_{\rm fs}^6 \nu^3 \nu'^3}{108 (2l+1) E_I^6}|\mathcal{M}(\nu)|^2,
\eeq
where the matrix element $\mathcal{M}(\nu)$ is given by Eq.~(B5) of Ref.~{\cite{Hirata_2photon}}, and $\nu' \equiv |\nu - \nu_{n1}|$, where $\nu_{n1}$ is the frequency of the Ly-$n$ transition. For $\nu < \nu_{n1}$, $d\Lambda_{nl}/d\nu$ is the rate of spontaneous two-photon decays from $nl$ per frequency interval. For $\nu > \nu_{n1}$, $d\Lambda_{nl}/d\nu \times f_{\nu'}$ (where $f_{\nu}$ is the photon occupation number at frequency $\nu$) is the rate of spontaneous Raman scatterings per frequency interval per atom initially in $nl$ (in the notation of Ref.~\cite{Hirata_2photon}, $d\Lambda_{nl}/d\nu = d K_{nl}/d\nu$ for $\nu > \nu_{n1}$). The function $d\Lambda_{nl}/d\nu$ is continuous across $\nu = \nu_{n1}$, where it vanishes.

We can now write the net rate of $nl \leftrightarrow 1s$ two-photon transitions per frequency interval per hydrogen atom, for which the highest energy photon has frequency $\nu < \nu_{n1}$:
\barr
\Delta_{nl}(\nu<\nu_{n1}) &=& \frac{d \Lambda_{nl}}{d \nu} \Big{[} x_{nl}(1 + f_{\nu'})(1 + f_{\nu}) \nonumber\\
&&- \frac{g_{nl}}{g_{1s}}x_{1s} f_{\nu'} f_{\nu}\Big{]}. \label{eq:Delta_nl}
\earr
For $\nu> \nu_{n1}$, the appropriate rate is that of Raman scattering events:
\barr
\Delta_{nl}(\nu>\nu_{n1}) &=& \frac{d \Lambda_{nl}}{d \nu}\Big{[} x_{nl}f_{\nu'}(1 + f_{\nu})\nonumber\\
&&- \frac{g_{nl}}{g_{1s}} x_{1s} (1 + f_{\nu'}) f_{\nu} \Big{]}, \label{eq:Delta_nl_Raman}
\earr
In both cases, we can assume that the photon occupation number for the low-energy photons is that of a blackbody, since the optical depth for two-photon absorption of the low-energy photons is tiny (for a discussion, see Ref.~\cite{Hirata_2photon}). We therefore set $f_{\nu'} = (\rme^{h \nu'/\Tr} - 1)^{-1}$. 
Moreover, the photon occupation number for frequencies $\nu > \nu_{\rm Ly \alpha}/2$ is much smaller than unity: $f_{\nu} \ll 1$. This means that we can neglect stimulated emission by the high-energy photons in Eqs.~(\ref{eq:Delta_nl}) and (\ref{eq:Delta_nl_Raman}).
Given these considerations, the net rate of two-photon transitions can be written in the following form, valid for both $\nu < \nu_{n1}$ and $\nu > \nu_{n1}$:
\barr
\Delta_{nl}(\nu) &=& \frac{d \Lambda_{nl}}{d \nu} \big{|}\rme^{h(\nu - \nu_{n1})/\Tr}- 1\big{|}^{-1} \nonumber\\
  &&\times \left[x_{nl}- \frac{g_{nl}}{g_{1s}} x_{1s}\rme^{h(\nu - \nu_{n1})/\Tr} f_{\nu} \right]. \label{eq:Delta_nl_general}
\earr

\subsection{Resonant scattering in Lyman-$\alpha$}\label{sec:rayleigh}

We now consider pure scattering events,
\begin{equation}
{\rm H(1s)} + \gamma \rightarrow {\rm H(1s)} + \gamma.
\end{equation}
In the low-frequency limit this corresponds to the familiar Rayleigh scattering phenomenon; the cross section however has resonances at the Lyman-series lines, which correspond to resonant Rayleigh scattering.  

Rayleigh scattering events conserve the photon frequency in the atom's rest frame. In the comoving frame (frame in which the CMB appears isotropic), the frequency of the scattered photon appears shifted due to the thermal motions of the scatterers. The frequencies of the incoming and outgoing photons are however statistically correlated. Mathematically, there is a definite probability distribution $p(\nu,\nu')$, such that $p(\nu,\nu')d\nu'$ is the probability that the outgoing photon has frequency in $[\nu', \nu' + d \nu']$ given that the incoming photon had frequency $\nu$, and this function generally depends on both $\nu$ and $\nu'$. For $\Tm \ll h \nu$, which is the case near the Lyman lines, the variance of the frequency shifts imparted by thermally moving atoms is given by:
\beq
\langle \delta \nu^2\rangle \equiv \int (\nu' - \nu)^2 p(\nu, \nu') d \nu' = \frac{2 \Tm}{m_{\rm H}c^2} \nu^2.\label{eq:dnu2-scat}
\eeq
The rate of injection of photons per frequency interval at frequency $\nu$, due to resonant Rayleigh scattering in Ly-$\alpha$, can be written in the general form (neglecting stimulated scatterings):
\beq
\Delta_{1s}(\nu) = x_{1s} \Big{[}\int  f_{\nu'} R(\nu', \nu) d \nu' - \int f_{\nu} R(\nu, \nu') d \nu'\Big{]}, \label{eq:Delta_diff}
\eeq
where $R(\nu, \nu') = \frac{d\Lambda_{1s}}{d \nu} p(\nu, \nu')$ is the differential rate of scatterings per hydrogen atom in the ground state, per unit frequency interval for both the incoming and outgoing photons (it has units of s$^{-1}$Hz$^{-2}$). The scattering kernel must respect detailed balance:
\beq
R(\nu, \nu')\rme^{- h \nu/\Tm} = R(\nu', \nu) \rme^{- h \nu'/\Tm}.
\eeq
To be fully general one should compute the scattering kernel from first principles. However, simplifications can be easily made in various regimes. 

Far from any resonance, the rate of redshifting due to the Hubble expansion is much larger than the rate of frequency diffusion due to scattering (see for example the discussion in Ref.~\cite{AGH10}). We can neglect Rayleigh scattering there, and set $\Delta_{1s}(\nu) = 0$.

Near Lyman resonances, we have 
\beq
\frac{d \Lambda_{1s}}{d \nu}(\nu \approx \nu_{n1}) \approx 3 A_{np,1s} p_{\rm sc}^n \phi_{V, n}(\nu), 
\eeq
where $p_{sc}^n = A_{np,1s}/\Gamma_{np}$ is the scattering probability in the Lyman-$n$ line (the complementary events being two-photon absorptions and two-photon photoionizations), and $\phi_{V,n}(\nu)$ is the Voigt profile for the Ly-$n$ line. In the Doppler core, we can approximate the partial redistribution induced by scattering events by a complete redistribution, i.e. approximate $p(\nu, \nu') \approx \phi_{V,n}(\nu) \approx \phi_{D, n}(\nu)$, where $\phi_D$ is the Doppler profile. This approximation is valid because  in the Doppler core, complete redistribution recovers the correct rms frequency shift during scattering events, Eq.~(\ref{eq:dnu2-scat}) (if one averages over the frequencies of absorbed photons). 

In the damping wings of Lyman resonances above Ly$\alpha$, the rate of scatterings is of the same order as the rate of two-photon absorptions. Each scattering event shifts the photon frequency by a very small amount compared to the width over which the radiation field varies ($\delta \nu_{\rm rms}/\nu \sim 2.5 \times 10^{-5}$). Partial redistribution is therefore essentially coherent in the comoving frame, i.e. $p(\nu, \nu') \approx \delta(\nu' - \nu)$, which implies $\Delta_{1s}(\nu) \approx 0$. For a more quantitative argument, see Ref.~\cite{AGH10}.

The only frequency regime where Rayleigh scattering affects the radiation field in a non-trivial way is in the damping wings of Ly$\alpha$. In this line, indeed, scattering events are much more frequent than two-photon absorption events (by a factor of $\sim 10^4$). Resonant scattering therefore leads to a significant diffusion in frequency. Because the frequency shifts are small compared to the width over which the radiation field varies, the integral scattering operator can be approximated by a second order differential operator -- a Fokker-Planck operator \cite{Rybicki06, Rybicki_DellAntonio, Hirata_Forbes, CS09c}. For the purpose of numerical implementation, the relevant properties are (i) the fact that this operator is nearly local (it only connects neighboring bins in frequency) (ii) it must respect detailed balance and (iii) the diffusion rate must be correct. We will explain our numerical method for the implentation of Lyman-$\alpha$ diffusion in Sec.~\ref{sec:numerical-diff}. 

We note that a number of analytic treatments of Lyman-$\alpha$ scattering in the recombination epoch have been proposed in the past \cite{Grachev89, Krolik1989, Krolik1990, Grachev_Dubrovich91, Rybicki_DellAntonio}.  However, since two-photon emission and absorption act on the same region of frequency space, and since both processes involve high optical depth, an accurate recombination history can only be obtained by considering all processes simultaneously.

\subsection{The radiative transfer equation}
 
The radiative transfer equation for the photon occupation number is:
\beq
\frac{\partial f_{\nu}}{\partial t} - H \nu \frac{\partial f_{\nu}}{\partial \nu} = \frac{c^3 n_{\rm H}}{8 \pi \nu^2} \left[\sum_{n\geq 2,l} \Delta_{nl}(\nu) + \Delta_{1s}(\nu) \right], \label{eq:rad_trans_2g}
\eeq
where the left-hand-side is the derivative of the photon occupation number along a photon trajectory in the expanding universe, and the prefactor on the right-hand-side converts the number of photons per unit frequency per hydrogen atom to the photon occupation number. 

\subsection{Inclusion in the effective multi-level atom rate equations}

\subsubsection{Formal two-photon decay rates}

As mentioned earlier, including two-photon decays from states with $n>2$ and Raman scattering events poses a double-counting problem. In principle, to avoid this double counting issue, one should discard ``1+1'' decays (or decays following an absorption event, which is like a Raman scattering event on resonance) altogether. If one were to pursue this idea, one should not consider the $p$ states at all anymore (as they are formally only intermediate states in two-photon processes), but consider all $s$ and $d$ states as ``interface states'' and allow for two-photon recombinations to the ground state. The two-photon $nl \leftrightarrow 1s$ transition rates would then become:
\beq
\dot{x}_{nl}\big{|}_{1s}^{(2 \gamma)} = - \dot{x}_{1s}\big{|}_{nl}^{(2 \gamma)} = x_{1s} \tilde{R}^{\rm total}_{1s,nl} - x_{nl}\tilde{R}^{\rm total}_{nl,1s}, \label{eq:dotxnl-2g-formal}
\eeq
where the formal transition rates are given by:
\beq
\tilde{R}^{\rm total}_{1s,nl} \equiv \int \frac{d \Lambda_{nl}}{d \nu}\frac{g_{nl}}{g_{1s}} \big{|}\rme^{h(\nu_{n1} - \nu)/\Tr}- 1\big{|}^{-1} f_{\nu} d \nu \label{eq:tildeR1snl-formal}\\
\eeq
and
\beq
\tilde{R}^{\rm total}_{nl,1s} \equiv  \int \frac{d \Lambda_{nl}}{d \nu} \big{|}\rme^{h(\nu - \nu_{n1})/\Tr}- 1\big{|}^{-1} d \nu, \label{eq:tildeRnl1s-formal}
\eeq
where the integrals run from $\nu_{n1}/2$ to $\nu_c$. In principle Eq.~(\ref{eq:dotxnl-2g-formal})--(\ref{eq:tildeRnl1s-formal}), can be included in a standard or effective multi-level atom code, provided one solves simultaneously for the radiation field, using the radiative transfer equation Eq.~(\ref{eq:rad_trans_2g}).

\subsubsection{Decomposition into ``1+1'' transitions and non-resonant contributions}

Two-photon decays from higher excited states constitute, however, a correction to the recombination history computed in the standard ``1+1'' picture, and we would like to implement it as such. We start by formally separating the integrals in Eqs.~(\ref{eq:tildeR1snl-formal}) and (\ref{eq:tildeRnl1s-formal}) in two contributions: the resonant pieces, for $\nu \approx \nu_{n'1}$, and a non-resonant piece, for frequencies far enough from any resonance. We therefore rewrite, formally:
\barr
\tilde{R}^{\rm total}_{1s,nl} &=& \sum_{n'} \tilde{R}_{1s,nl}^{(n'p)} + \tilde{R}_{1s,nl} {\rm ~~and}\nonumber\\
\tilde{R}^{\rm total}_{nl,1s} &=& \sum_{n'} \tilde{R}_{nl,1s}^{(n'p)} + \tilde{R}_{nl,1s},                         
\earr  
where the resonant contributions $\tilde{R}_{1s,nl}^{(n'p)}$ and $\tilde{R}_{nl,1s}^{(n'p)}$ are defined in a similar manner as in Eqs.~(\ref{eq:tildeR1snl-formal}) and (\ref{eq:tildeRnl1s-formal}), but with the integration being carried over a narrow range $\Delta \nu$ near $\nu_{n'1}$, and $\tilde{R}_{1s,nl}$ and $\tilde{R}_{nl,1s}$ are the non-resonant pieces required to complete the total rates. So far the separation is just formal and we have not made any approximation. 

\subsubsection{``1+1'' Resonant contribution}

We now notice that near a resonance $\nu \approx \nu_{n'1}$, the two-photon differential decay rate $d \Lambda_{nl}/d \nu$ takes on the following form (if $n > n'$):
\barr
\frac{d \Lambda_{nl}}{d \nu}\Big{|}_{\nu \approx \nu_{n'1}} &\approx& \frac{1}{4 \pi^2}\frac{A_{nl,n'p}A_{n'p,1s}}{(\nu - \nu_{n'1})^2 + (\Gamma_{n'p}/4\pi)^2}\nonumber\\
&=& A_{nl,n'p} \frac{A_{n'p,1s}}{\Gamma_{n'p}}  \phi_L(\nu-\nu_{n'1}; \Gamma_{n'p}),\;\;\;\; \label{eq:dLnldnu-resonance}
\earr  
where $\Gamma_{n'p}$ is the total inverse lifetime of the state $n'p$, and the Lorentzian profile is given by
\beq
\phi_L(\Delta \nu; \Gamma) \equiv \frac{\Gamma/(4 \pi^2)}{\Delta \nu^2 + (\Gamma/4\pi)^2}.
\eeq
For $n<n'$, the first coefficient in Eq.~(\ref{eq:dLnldnu-resonance}) should be $ g_{n'p}/g_{nl} \times A_{n'p,nl}$ instead of $A_{nl,n'p}$. When accounting for the thermal motions of atoms, the Lorentzian profile should be replaced by a Voigt profile. We can now approximate the resonant pieces with the following expressions, valid for both $n<n'$ and $n>n'$:
\beq
\tilde{R}_{1s,nl}^{(n'p)} \approx 3 A_{n'p,1s} \overline{f}_{\nu_{n'1}} \frac{R_{n'p,nl}}{\Gamma_{n'p}}
\label{eq:1snl-resonance}
\eeq
and
\beq
\tilde{R}_{nl,1s}^{(n'p)} \approx R_{nl,n'p} \frac{A_{n'p,1s}}{\Gamma_{n'p}},\label{eq:nl1s-resonance}
\eeq
where $\overline{f}_{\nu_{n'1}}$ is the photon occupation number averaged over the Voigt profile near the resonance $\nu \approx \nu_{n'1}$. Eqs.~(\ref{eq:1snl-resonance}) and (\ref{eq:nl1s-resonance}) are exactly what one would obtain in the ``1+1'' picture after ``factoring out'' the $p$ states (with a procedure similar to what is used to get rid of the ``interior'' states in the EMLA method). Having these resonant rates is exactly equivalent to having optically thin one-photon transitions between the $nl$ and $n'p$ states, with rates $R_{nl,n'p}(\Tr)$ and $R_{n'p,nl}(\Tr)$, and optically thick Lyman transitions, with net rate:
\beq
\dot{x}_{n'p}\big{|}_{1s} = - \dot{x}_{1s}\big{|}_{n'p} = A_{n'p,1s}\left(3 x_{1s} \overline{f}_{\nu_{n'p}} - x_{n'p}\right). \label{eq:dotxnp-resonant}
\eeq 
To obtain the net decay rates in the Lyman transitions, one then needs to solve for the radiation field in the immediate vicinity of Lyman resonances. If the frequency region for which two-photon transitions are considered as ``resonant'' is narrow enough, this can be done in the Sobolev approximation. Indeed, all the relevant conditions are met (see also discussion in Ref.~\cite{Hirata_2photon}; for more details on the Sobolev approximation, see for example Ref.~\cite{AGH10}): 

First, the two-photon absorption and emission profiles can both be approximated by the same resonance profile Eq.~(\ref{eq:dLnldnu-resonance}). This relies on the assumption that the blackbody radiation field varies little across the ``resonant'' region, and requires for its width to satisfy $\Delta \nu \ll \Tr/h$. 

Secondly, we argued in Sec.~\ref{sec:rayleigh} that one could assume complete frequency redistribution for resonant scattering near the Doppler core of Lyman resonances. The ``resonant'' region should therefore not exceed a few Doppler widths.

Finally, if we consider regions in frequency narrow enough around the resonances, we can use the steady-state approximation. This requires $\Delta \nu/\nu \ll 1$.

We can see that considering the ``resonant'' region around each Lyman resonance to be a few Doppler widths wide meets all the requirements. 

An additional assumption required here is that excited states are near Boltzmann equilibrium, which is very accurate at redshifts for which two-photon processes are important.  In the Sobolev approximation, and in the limit of large Sobolev optical depth, Eq.~(\ref{eq:dotxnp-resonant}) becomes the standard Lyman decay rate Eq.~(\ref{eq:dotxnp feedback}), where $f_{np}^+$ is the photon occupation number incoming on the resonance, preprocessed by two-photon processes and diffusion in the blue damping wing of the line. 

The Sobolev approximation is probably the least accurate for Ly$\alpha$ decays where partial redistribution due to resonant scattering is important. However, the large optical depth to two-photon absorptions in the Lyman-$\alpha$ blue damping wing, in conjunction with frequent scatterings, drive the radiation field to the equilibrium value $f_{\nu} = x_{n'p}/(3 x_{1s}) \rme^{-h(\nu - \nu_{n'1})/\Tm}$ over several Doppler widths (of the order of 40 Doppler widths, see Ref.~\cite{AGH10}). As a consequence the net decay rate in the core of the resonance is very small anyway. We checked that in the presence of two-photon transitions and frequency diffusion, even setting $\dot{x}_{n'p}\big{|}_{1s} = 0$ instead of the expression given by Eq.~(\ref{eq:dotxnp-resonant}) leads to relative changes to the recombination history of at most $7 \times 10^{-4}$. Given that frequency diffsion leads to corrections of a few percent at most to the decay rate in Ly$\alpha$ when radiative transfer is treated carefully even at the line center \cite{Hirata_Forbes}, we can be confident that using the Sobolev approximation for the resonant contributions of two-photon decays is accurate to better than $10^{-4}$.

\subsubsection{``Pure two-photon'' non-resonant contribution}

In the previous section we discussed how two-photon decays within a few Doppler widths of Lyman resonances can in fact be accounted for in the standard ``1+1'' picture. To evaluate the non-resonant pieces, $\tilde{R}_{1s,nl}$ and $\tilde{R}_{nl,1s}$, we need to solve the radiative transfer equation, Eq.~(\ref{eq:rad_trans_2g}), to obtain the photon occupation number. The subject of Sec.~\ref{sec:two-photon numerical} is to describe our numerical method of solution.

Note that choosing the ``resonant'' regions to be a few Doppler widths has an additional advantage. Since a Doppler width is $\sim 10^3$ times wider than the natural width of Lyman lines, it is not necessary to account for the pole displacements in the computation of the differential two-photon decay rates in the non-resonant region. In addition, the fraction of two-photon decays that are considered non-resonant will be small (of the order of $\Gamma_{np}/(4\pi^2)/\Delta \nu$, where $\Delta \nu$ is the width of the ``resonant'' region). For $\Delta \nu$ of a few Doppler widths, this fraction is $\sim 10^{-4}$. This means that the ``pure'' two-photon decay rates $\tilde{R}_{nl, 1s}$ are much smaller than the total inverse lifetime of the $nl$ state, $\Gamma_{nl}$, which is required to simplify the effective MLA model to an effective four-level atom model as we discussed in Sec.\ref{sec:E4LA}.

As a final note, we want to emphasize why the final result is independent of the exact boundary between ``resonant'' and ``non-resonant'' regions, so long as the resonant regions are a few Doppler widths wide. If one were to increase the width of the ``resonant'' region, then the ``pure'' two-photon transition rates $\tilde{R}_{nl,1s}$ and $\tilde{R}_{1s,nl}$ would decrease, mainly because of the change of the integration region in the blue wings of the resonance -- in the red wing, the radiation field has reached near equilibrium with the line and the net rate of decays immediately blueward of line center is very small anyway. This decrease would be nearly exactly compensated by the increase of what is considered as ``1+1'' decays, as the photon occupation number incoming on the Lyman resonances, $f_{np}^+$, would be decreased due to the smaller optical depth due to ``pure'' two-photon absorptions in the blue wing. Hirata (2008) checked the independence of the result form the exact value chosen for the width of the ``resonant'' region, and found that even changing this width by a factor of 9 lead to relative changes of at most $4 \times 10^{-4}$ in the recombination history.

\section{Numerical solution of the radiative transfer equation} \label{sec:two-photon numerical}

\subsection{Discretization of the radiative transfer equation}\label{sec:numerical-diff}

To solve the radiative transfer equation [Eq.~(\ref{eq:rad_trans_2g})] numerically in the ``non-resonant'' frequency region, we follow the method of Hirata (2008), and extend it to also account for frequency diffusion.

We will consider the radiation field in the vicinity of $N$ frequency ``spikes'' $\nu_b$, for  $b = 1,2, ...~N$. Each spike has an associated width $\Delta \nu_b$ (which is just the separation between consecutive spikes if they are linearly spaced for example). 

We use the discretized differential two-photon rate:
\beq
\frac{d \Lambda_{nl}}{d \nu}\Big{|}_{\rm used} = \sum_b A_{nl,b} \delta_{\epsilon}(\nu-\nu_b), \label{eq:dLnl_db used}
\eeq 
where we use the coefficients
\beq
A_{nl,b} \equiv \int_{\Delta \nu_b} \frac{d \Lambda_{nl}}{d \nu} d\nu,
\eeq
where the integral is carried over the frequency region associated with the spike $\Delta \nu_b$. The function $\delta_{\epsilon}(\nu - \nu_b)$ in Eq.~(\ref{eq:dLnl_db used}) should be understood as a sharp profile centered at $\nu_b$, which integrates to unity, and has support in $[\nu_b - \epsilon, \nu_b + \epsilon]$. The solution we derive is in the limit $\epsilon \rightarrow 0$, for which $\delta_{\epsilon} \rightarrow \delta$, the Dirac delta function. Doing so, we are simply approximating the optical depth as concentrated in discrete frequencies instead of being a smooth function.

The main new contribution of the present work is the discretization method for the scattering operator. We use the discretized scattering kernel
\beq
R(\nu, \nu')\Big{|}_{\rm used} = \sum_{b, b'} R_{b,b'} \delta_{\epsilon}(\nu - \nu_b) \delta_{\epsilon}(\nu' - \nu_{b'}).  
\eeq
We enforce detailed balance:
\beq
R_{b,b'}\rme^{- h \nu_b/\Tm} = R_{b',b} \rme^{- h \nu_{b'}/\Tm}. \label{eq:diffusion-db}
\eeq
We moreover use the diffusion approximation for resonant scattering. This allows us to assume that the numerical scattering kernel $R_{b,b'}$ is non-vanishing only for neighboring bins, $b' = b \pm 1$. In order to obtain the correct diffusion rate, we set
\barr
&&(\nu_{b+1} - \nu_b)^2 R_{b,b+1}+ (\nu_{b-1} - \nu_b)^2 R_{b, b-1}  \nonumber \\
&&= 3 \frac{A_{2p,1s}^2}{4 \pi^2 (\nu - \nu_{\rm Ly \alpha})^2} \Delta \nu_b \frac{2 \Tm}{m_{\rm H} c^2} \nu_{\rm Ly \alpha}^2,  \label{eq:diffusion-rate}
\earr
where we used the damping wing approximation for the absorption profile (and approximate $\nu_b^2 \approx \nu_{\Ly \alpha}^2$ in the multiplicative factor).

As boundary conditions, we assume a vanishing photon flux due to diffusion at the boundaries of our domain, i.e., formally, $R_{1,0} = R_{N, N +1} = 0$ (in fact we set these conditions at the boundaries of the diffusion domain, smaller than the entire frequency domain considered). Using Eq.~(\ref{eq:diffusion-rate}), we then obtain $R_{1,2}$ and $R_{N, N - 1}$. Using iteratively Eqs.~(\ref{eq:diffusion-db}) and (\ref{eq:diffusion-rate}), we can then obtain all the coefficients of the numerical diffusion kernel, starting from the boundaries, and up to line center. Denoting $b_1$ the highest bin below Ly-$\alpha$ and $b_1+1$ the first bin above Ly-$\alpha$, we obtain all coefficients up to $R_{b_{\rm Ly \alpha}, b_1}$ on the red side of Ly-$\alpha$, and up to $R_{b_{\rm Ly \alpha}, b_1+1}$ on the blue side (we do not follow the radiation field at the central bin $b_{\Ly \alpha}$ but can still define these coefficients). Note that with this method we cannot ensure that the diffusion rate at the central bin is correct. However, the exact value of the diffusion rate at line center does not matter, as long as it is high enough to ensure that the photon occupation number reaches the equilibrium spectrum $f_{\nu} \propto \rme^{- h \nu/\Tm}$.

\subsection{Solution of the discretized radiative transfer equation}\label{sec:rad trans sol}

To simplify the notation, we define the following rate coefficients:
\barr
R_{nl,b} &\equiv& \frac{d \Lambda_{nl}}{d \nu}\Big{|}_{\nu_b} \big{|}\rme^{h(\nu_b - \nu_{n1})/\Tr}- 1\big{|}^{-1} \Delta \nu_b
{\rm~~and}\nonumber\\
R_{b,nl} &\equiv& \frac{g_{nl}}{g_{1s}}\rme^{h(\nu_b - \nu_{n1})/\Tr}R_{nl,b}.
\earr 
This coefficients can be thought of as transition rates between bound states and a set of ``virtual'' levels with associated energies $E_b = h \nu_b$ \cite{Hirata_2photon}.

We define the total Sobolev optical depth in the $b$-th frequency spike:
\beq
\Delta \tau_b \equiv \frac{c^3 n_{\rm H} x_{1s}}{8 \pi \nu_b^3 H}\left(\sum_{nl}R_{b,nl} + \sum_{b' = b\pm1}R_{b,b'} \right)
\eeq
We also define the average photon occupation number near $\nu_b$:
\beq
\overline{f}_{\nu_b} \equiv \int_{\nu_b-\epsilon}^{\nu_b + \epsilon} \delta_{\epsilon}(\nu - \nu_b) f_{\nu} d \nu.
\eeq
Finally, we define the equilibrium photon occupation number at the $b$-th frequency spike:
\beq
f_{\nu_{b}}^{\rm eq} \equiv \frac{\sum_{nl} x_{nl} R_{nl,b} + x_{1s} \sum_{b' } \overline{f}_{\nu_{b'}} R_{b',b} }{ x_{1s}\left(\sum_{nl} R_{b,nl} + \sum_{b' }  R_{b,b'}\right) } \label{eq:fnub-eq}
\eeq
In the vicinity of $\nu_b$, the discretized radiative transfer equation becomes:
\beq
\frac{1}{H \nu_b} \frac{\partial f_{\nu}}{\partial t} - \frac{\partial f_{\nu}}{\partial \nu} = \Delta \tau_b \delta_{\epsilon}(\nu - \nu_b)\left[ f_{\nu_b}^{\rm eq} - f_{\nu}\right].
\eeq
In the limit that the support of the delta function becomes vanishingly small, $\epsilon \rightarrow 0$, the discretized radiative transfer equation can be solved in the steady-state approximation, and one can neglect the time derivative. This is similar to the commonly used Sobolev approximation, except that we are now making this approximation in the vicinity of an artificially introduced spike (as opposed to a true resonance line), for the purposes of numerical resolution. Another conceptual difference is that the equilibrium photon occupation number also depends on the averaged value of the radiation field at neighboring bins, because of frequency diffusion. Given the photon occupation number at the blue edge of the $b$-th spike, $f_{\nu_b + \epsilon}$, this equation has a well known solution $f_{\nu}$. The quantities of interest for us are the photon occupation number at the red edge of the spike $f_{\nu_b - \epsilon}$ and the average photon occupation number in the spike $\overline{f}_{\nu_b}$. They are given by the following expressions (for a derivation, see for example Refs.~\cite{AGH10} and \cite{Hirata_2photon}):
\beq
f_{\nu_b - \epsilon} =  f_{\nu_b + \epsilon} \rme^{- \Delta \tau_b} + f_{\nu_b}^{\rm eq} \left(1 - \rme^{- \Delta \tau_b}\right) , \label{eq:fnub-epsilon}
\eeq
and
\beq
\overline{f}_{\nu_b} =   \Pi_b f_{\nu_b + \epsilon} + (1 - \Pi_b)  f_{\nu_b}^{\rm eq} , \label{eq:fnub-average}
\eeq
where $\Pi_b$ is the Sobolev escape probability from the $b$-th spike:
\beq
\Pi_b \equiv \frac{1 - \rme^{- \Delta \tau_b}}{\Delta \tau_b}.
\eeq
We now use the variables
\beq
x_b \equiv x_{1s} \overline{f}_{\nu_b}.
\eeq
As explained in Ref.~\cite{Hirata_2photon}, $x_b$ can be interpreted as the population of the virtual level $b$. One should however keep in mind that this is is just a convenient rewording for the radiation field intensity.

Using the definition of $f_{\nu_b}^{\rm eq}$, Eq.~(\ref{eq:fnub-eq}), we can rewrite Eq.~(\ref{eq:fnub-average}) in the form:
\beq
T_{b,b} x_b =  \sum_{nl} x_{nl} R_{nl,b} +  \sum_{b' = b\pm 1} x_{b'} R_{b',b}  + s_b, \label{eq:fnub-linear-eq}
\eeq
where we have defined:
\barr
T_{b,b} &\equiv& \frac{1}{1 - \Pi_b} \left(\sum_{nl} R_{b,nl} + \sum_{b' =b\pm1}  R_{b,b'}\right) {\rm~~and}\nonumber\\
s_b &\equiv& \Pi_b x_{1s} f_{\nu_b + \epsilon} T_{b,b}.
\earr

We only follow two-photon decays in the damping wings of resonances, but we should still account for frequency diffusion between line center and the neighboring bins. At the Lyman-$\alpha$ line center, the radiation field is in equilibrium with the $2p/1s$ ratio: $f_{\nu_{\rm Ly \alpha}} = x_{2p}/(3 x_{1s})$. If $b_1$ is the highest frequency bin below Ly-$\alpha$, (and $b_1+1$ is the first bin above Ly-$\alpha$), we therefore define the transition rates with $2p$:
\beq
R_{2p,b_1} = \frac{1}{3}R_{b_{\rm Ly \alpha}, b_1} {\rm~~and~~} R_{2p,b_1+1} = \frac{1}{3}R_{b_{\rm Ly \alpha}, b_1+1}.
\eeq
Provided that we set $R_{b_1, b_1+1} = R_{b_1+1, b_1} = 0$, Eq.~(\ref{eq:fnub-linear-eq}) remains valid for $b = b_1, b_1 +1$.
Adding these transitions with the central frequency bin will ensure that the photon occupation number is driven to its equilibrium value near line center, $f_{\nu}^{\rm eq} = x_{2p}/(3 x_{1s}) \rme^{-h(\nu - \nu_{\rm Ly \alpha})/\Tm}$. 

We now use Eq.~(\ref{eq:XK-effective}) for the populations $x_{nl}$ with $n \geq 3$. We define the coefficients, for $i = 2s, 2p$:
\beq
T_{b,i} \equiv -R_{i,b} - \sum_{n\geq3, l} \frac{g_{nl}}{g_i} \rme^{-E_{n2}/\Tr} P_{nl}^i(\Tr) R_{nl,b}. \label{eq:Tbi}
\eeq
We define the new source vector:
\beq
S_b \equiv s_b + x_e^2\sum_{n \geq 3, l} \frac{g_{nl}}{g_e} \rme^{-E_n/\Tr} P_{nl}^e(\Tr)  R_{nl,b},
\eeq
We also define the coefficients
\barr
T_{b, b\pm1} \equiv - R_{b\pm1, b}.
\earr
The discretized radiative transfer equation then takes the final form:
\beq
T_{b,2s} x_{2s} + T_{b,2p} x_{2p} +  \sum_{b' = b-1}^{b+1} T_{b,b'} x_{b'} = S_b. \label{eq:rad-trans-final}
\eeq

\subsection{Populations of the excited states}

Given the radiation field, we can now compute the two-photon transition rates. Using Eqs.~(\ref{eq:tildeR1snl-formal}) and (\ref{eq:tildeRnl1s-formal}) limited to the ``non-resonant'' frequency region, we obtain, after discretization:
\beq
\tilde{R}_{nl,1s} = \sum_{b}R_{nl,b} {\rm~~and~~}
\tilde{R}_{1s,nl} = \sum_b \overline{f}_{\nu_b}  R_{b,nl}. 
\eeq  
The effective transition rates from the $i = 2s$ and $2p$ states to the ground state are therefore, according to the discussion in Sec.~\ref{sec:E4LA}, and using the definition of $T_{b,i}$ Eq.~(\ref{eq:Tbi}):
\beq
\mathcal{\tilde{R}}_{i,1s} = - \sum_b T_{b,i} + \sum_{n} \frac{g_{np}}{g_i} R_{\textrm{Ly}n}\rme^{- E_{n2}/\Tr} P_{np}^{i}(\Tr),
\eeq
where the first term accounts for two-photon transitions and the second term for escape from Lyman lines (it is understood that $P_{2p}^{2s} = 0$ and $P_{2p}^{2p} = 1$). The effective transition rate for the reverse process is given by:
\beq
\mathcal{\tilde{R}}_{1s,i} = - \sum_b T_{i,b} \overline{f}_{\nu_b}
+ 3 \sum_{n} R_{\textrm{Ly}n}P_{np}^{i}(\Tr) f_{np}^+,
\eeq
where we have defined the coefficients, for $i = 2s, 2p$:
\beq
T_{i,b} \equiv \frac{g_i}{g_{1s}}\rme^{h(\nu_b - \nu_{21})/\Tr} T_{b,i}.
\eeq
For a given radiation field and free electron fraction, we can now obtain an equation for the populations of the excted states $2s, 2p$. We do so in the steady-state approximation, i.e. setting $\dot{x}_{i} = 0$ for $i = 2s, 2p$ in Eqs.~(\ref{eq:x2sdot}) and (\ref{eq:x2pdot}). We first define the $2\times 2$ matrix of elements
\barr
T_{i,i} &\equiv& \mathcal{B}_i + \mathcal{R}_{i,j} + \mathcal{\tilde{R}}_{i,1s} {\rm~~and}\nonumber \\
T_{i,j} & \equiv& - \mathcal{R}_{i,j}.
\earr
We also define the source vector of elements
\beq
S_i \equiv n_{\rm H} x_e^2 \mathcal{A}_i + 3 \sum_{n\geq 2}  R_{\textrm{Ly}n}  P_{np}^{i}(\Tr) f_{np}^+.
\eeq
The steady-state equation for each state $i$ translates into the linear equation:
\beq
\sum_{j=2s,2p}T_{i,j} x_j + \sum_b T_{i,b} x_b = S_i, \ \ \ i = 2s, 2p. \label{eq:pop-final}
\eeq

\subsection{Evolution of the coupled system of level populations and radiation field}

\begin{figure} 
\includegraphics[width = 75mm]{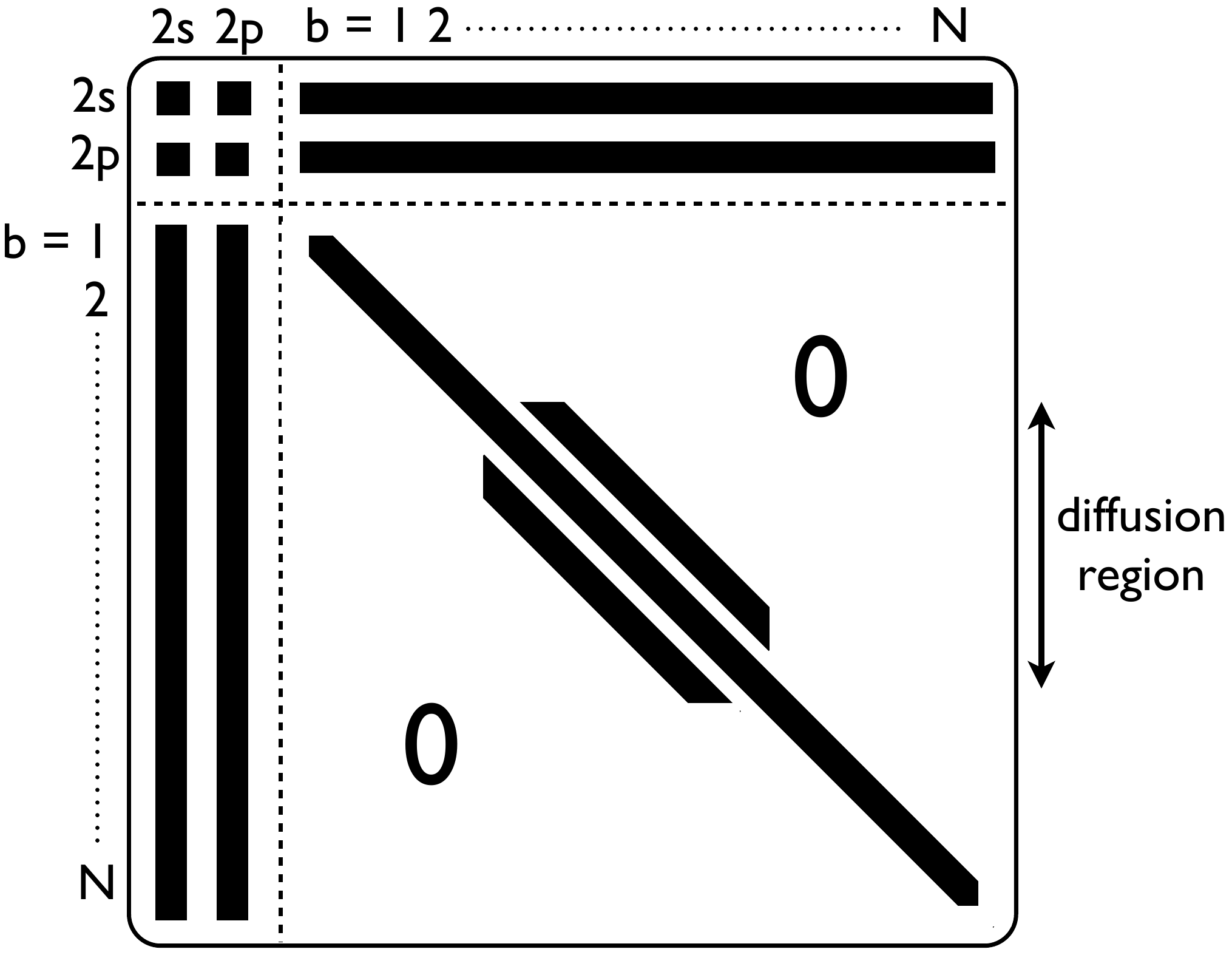}
\caption{Sparsity pattern of the linear system solved for evolving simultaneously the level populations and the radiation field, in the presence of two-photon transtions and frequency diffusion.}
\label{fig:matrix}
\end{figure}

We now have all the necessary pieces to evolve simultaneously the level populations and the radiation field, and compute the free electron fraction. In this section we summarize the procedure and recall the main equations.
We start with an initially thermal radiation field, $f_{\nu} = \rme^{-h\nu/\Tr}$. At each time step, we do the following computations:
\newcounter{XL}
\begin{list}{\arabic{XL}.}{\usecounter{XL}}
\item
We obtain the photon occupation number incoming on each bin $b$ assuming free streaming between frequency spikes:
\beq
f_{\nu_b+\epsilon}(z) = f_{\nu_{b+1}-\epsilon}\left(z' = (1+z) \frac{E_{b+1}}{E_b} - 1\right).
\eeq
We also obtain in the same way the incoming photon occupation number at the Ly-$n$ transitions, $f_{np}^+$.
\item\label{it:ii}
We solve for the populations of the $2s$ and $2p$ states and the average photon occupation number at each frequency spike $\overline{f}_{\nu_b} = x_b/x_{1s}$ \emph{simultaneously} by solving the coupled linear system given by Eqs.~(\ref{eq:rad-trans-final}) and (\ref{eq:pop-final}). Even with a large number of bins for the radiation field ($N = 311$ in our fiducial case), this system is easily solved because the matrix of coefficients $T_{b,b'}$ is triadiagonal and the overall system has the particular sparsity pattern shown in Fig.~\ref{fig:matrix}. Such a sparse system can be solved in $\mathcal{O}(N)$ operations (specifically, we can solve the system in $\sim 16 N$ operations).
\item
We update the photon occupation number at the red side of each spike, $f_{\nu_b - \epsilon}$, using Eq.~(\ref{eq:fnub-epsilon}). At the red side of Lyman resonances, we use $f_{np}^- = x_{np}/(3 x_{1s})$, valid in the optically thick limit, where $x_{np}$ is given by Eq.~(\ref{eq:XK-effective}) for $n \geq 3$.
\item
After step \#\ref{it:ii}, we can obtain the function $\dot{x}_e(z, x_e)$ through Eq.~(\ref{eq:xedot EFLA}) or\footnote{Eq.~(\ref{eq:xedot EFLA}) contains near exact cancellations but Eq.~(\ref{eq:x1sdot EFLA}) contains a large number of terms for which numerical roundoff errors can add up. We checked that both equations give the same result within numerical roundoff errors. We use Eq.~(\ref{eq:xedot EFLA}) in the final code simply because it is more compact.} (\ref{eq:x1sdot EFLA}). This allows us to evolve the free electron fraction to the next timestep. \end{list}

\subsection{Implementation, convergence tests and results}\label{sec:results}

We evolve the free electron fraction during hydrogen recombination in several phases. We use even steps in $\ln a$ (where $a = (1+z)^{-1}$ is the scale factor), with $\Delta \ln a = 8.5 \times 10^{-5}$. We describe our ODE integrator in Appendix \ref{app:ODE}. 

\begin{list}{$\bullet$}{}
\item We checked that hydrogen and helium recombination never overlap and can be followed separately (to an accuracy of a few times $10^{-4}$). We therefore start computing the hydrogen recombination history once helium is completely recombined. Quantitatively, we start hydrogen recombination once the fractional abundance of He$^+$ ions is less than $10^{-4}$ relative to hydrogen. If this criterion is met earlier than $z=1650$ we only switch on the hydrogen recombination computation at $z = 1650$. We checked that at this redshift the exact free electron fraction differs from the Saha equilibrium value by no more than a few times $10^{-4}$ anyway.

\item In the first phases of hydrogen recombination, we use the post-Saha expansion described in Appendix \ref{app:post-Saha}. We do so as long as the free electron fraction differs from the Saha value by $\Delta x_e < 5\times 10^{-5}$. We checked that explicitly integrating the ODE for $x_e$ instead (with a much smaller timestep as the ODE is stiff at early times) leads to maximum changes of $\Delta x_e/x_e \lesssim 3\times 10^{-4}$.

\item From then on and until $ z = 700$ we solve simultaneously for the level populations and radiative transfer with two photon processes and diffusion as described in this section.
\item For $z < 700$, we use the simple EMLA equations, with simple decay rates from $2s$ and $2p$ only (i.e. not accounting for higher order Lyman lines and radiative transfer effects). We checked that moving the last switch to $z = 400$ instead of 700 leads to maximum changes $|\Delta x_e|/x_e < 10^{-4}$.

\item For the matter temperature evolution, we use the asymptotic solution of Hirata \cite{Hirata_2photon} [it can be obtained by setting $\dot{T}_{\rm m} = - H\Tm$ in Eq.~(\ref{eq:Tmdot})] as long as $1-\Tm/\Tr < 5\times 10^{-4}$. Depending on cosmology, this corresponds to $750 < z < 950$. After that we switch to solving for $x_e$ and $\Tm$ simultaneously by using Eq.~(\ref{eq:Tmdot}).

\end{list}

All the checks mentioned above were made for a wide range of cosmological parameters.

Our fiducial parameters for the numerical solution of radiative transfer are $N = 311$ frequency bins extending from $\nu_{\Ly \alpha}/2$ to $\nu_{\Ly \gamma}$, and a diffusion region with 80 bins extending to $\Delta \nu/\nu_{\rm Ly \alpha} = \pm 1.7 \times 10^{-2}$. The minimal spacing between bins is $\min[\ln(\nu_{b+1}/\nu_b)] = 8.5 \times 10^{-5}$, which sets the largest step in $\ln a$ that we can take (this is also half of the width $\Delta \nu/\nu$ of the ``resonant'' region around Ly$\alpha$). We checked (for the fiducial cosmology only) that reducing the diffusion region to $\Delta \nu/\nu_{\rm Ly \alpha} = \pm 1\times 10^{-2}$ leads to changes $|\Delta x_e|/x_e < 6 \times 10^{-6}$. Reducing the diffusion region to $\Delta \nu/\nu_{\rm Ly \alpha} = \pm 5\times 10^{-3}$ leads to changes $|\Delta x_e|/x_e < 4 \times 10^{-5}$. We checked that using a 10 times finer frequency grid in the diffusion region (and a 10 times smaller timestep) leads to maximum changes $|\Delta x_e|/x_e \approx 1.5 \times 10^{-4}$ at $z \approx 900$. 

We are therefore confident that our numerical treatment is converged at the level of a few parts in $10^{4}$.

\begin{figure} 
\includegraphics[width = 85mm]{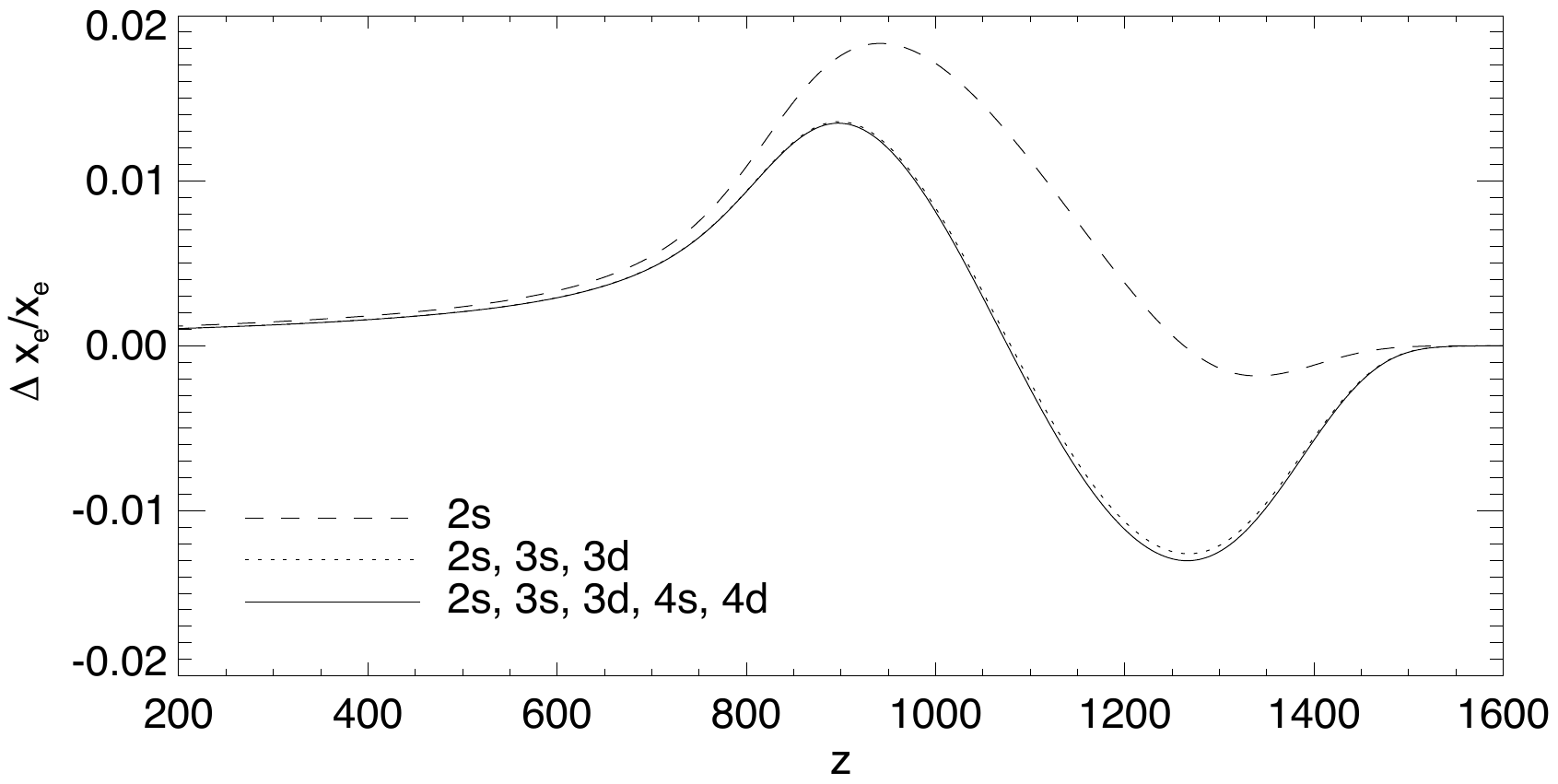}
\caption{Changes in the recombination history when including two-photon decays and Raman scattering (no diffusion), compared to our ``base'' model. The line labeled '$2s$' shows the changes in $x_e$ when one properly accounts for stimulated $2s\rightarrow 1s$ decays, as well as absorptions of distortion photons and Raman scattering form $2s$. The other lines show the cumlative correction when adding two-photon transitions from higher levels.}
\label{fig:two_photon}
\end{figure}

\begin{figure} 
\includegraphics[width = 85mm]{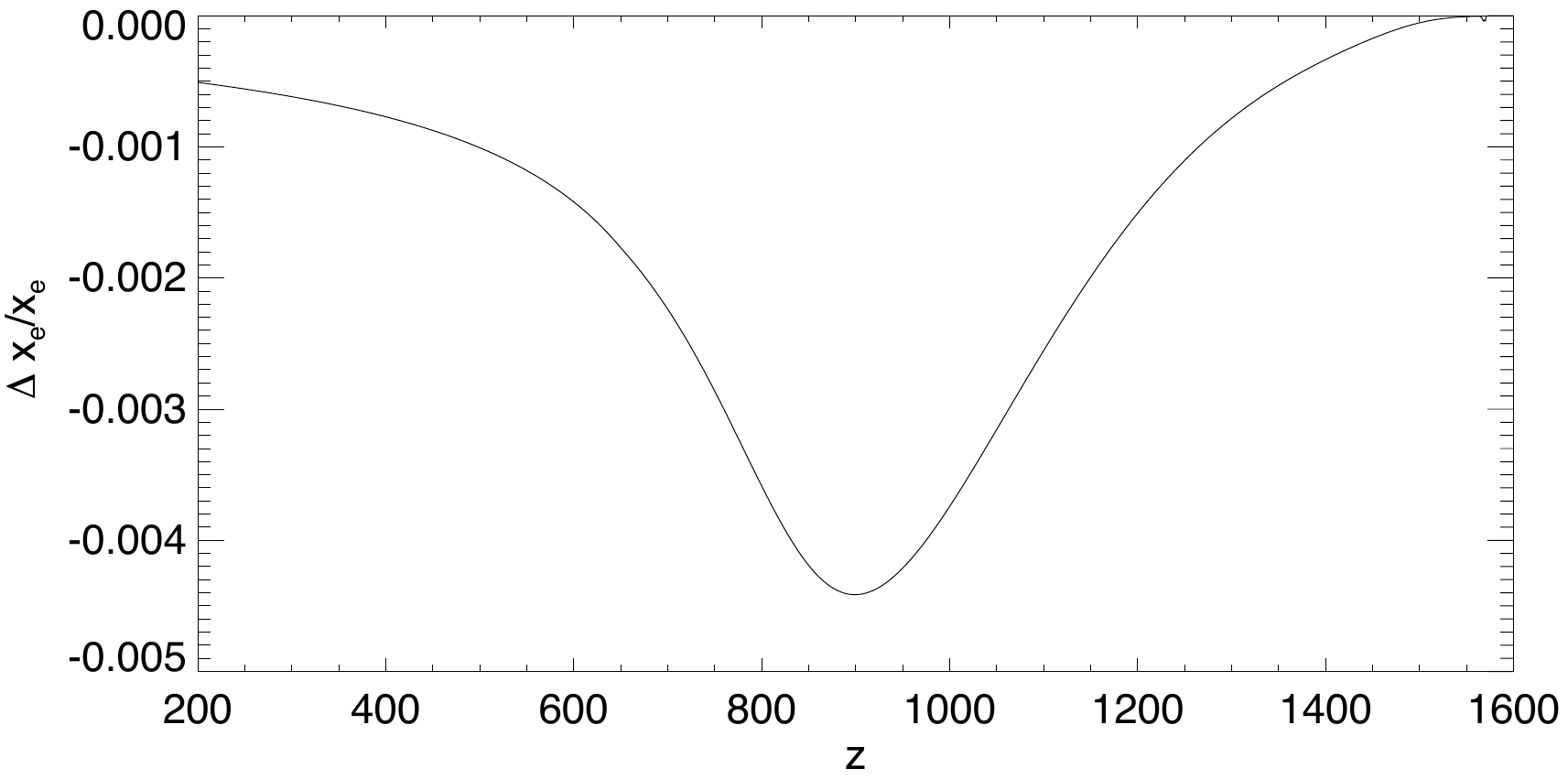}
\caption{Changes in the recombination history when including frequency diffusion in Lyman-$\alpha$, compared to a model with two-photon transitions but no diffusion.}
\label{fig:diffusion}
\end{figure}

We show in Fig.~\ref{fig:two_photon} the changes in the recombination history due to two-photon processes. We find that including two-photon transitions from the initial states $2s, 3s, 3d, 4s$ and $4d$ is sufficient for the level of accuracy required -- we checked that including two-photon transitions from $5s$ and $5d$ leads to a maximum change $\Delta x_e/x_e \sim 8 \times 10^{-5}$ at $z\sim 1200$ and can therefore be neglected. The effect of frequency diffusion in the Ly$\alpha$ line is shown in Fig.~\ref{fig:diffusion}.

\begin{figure} 
\includegraphics[width = 85mm]{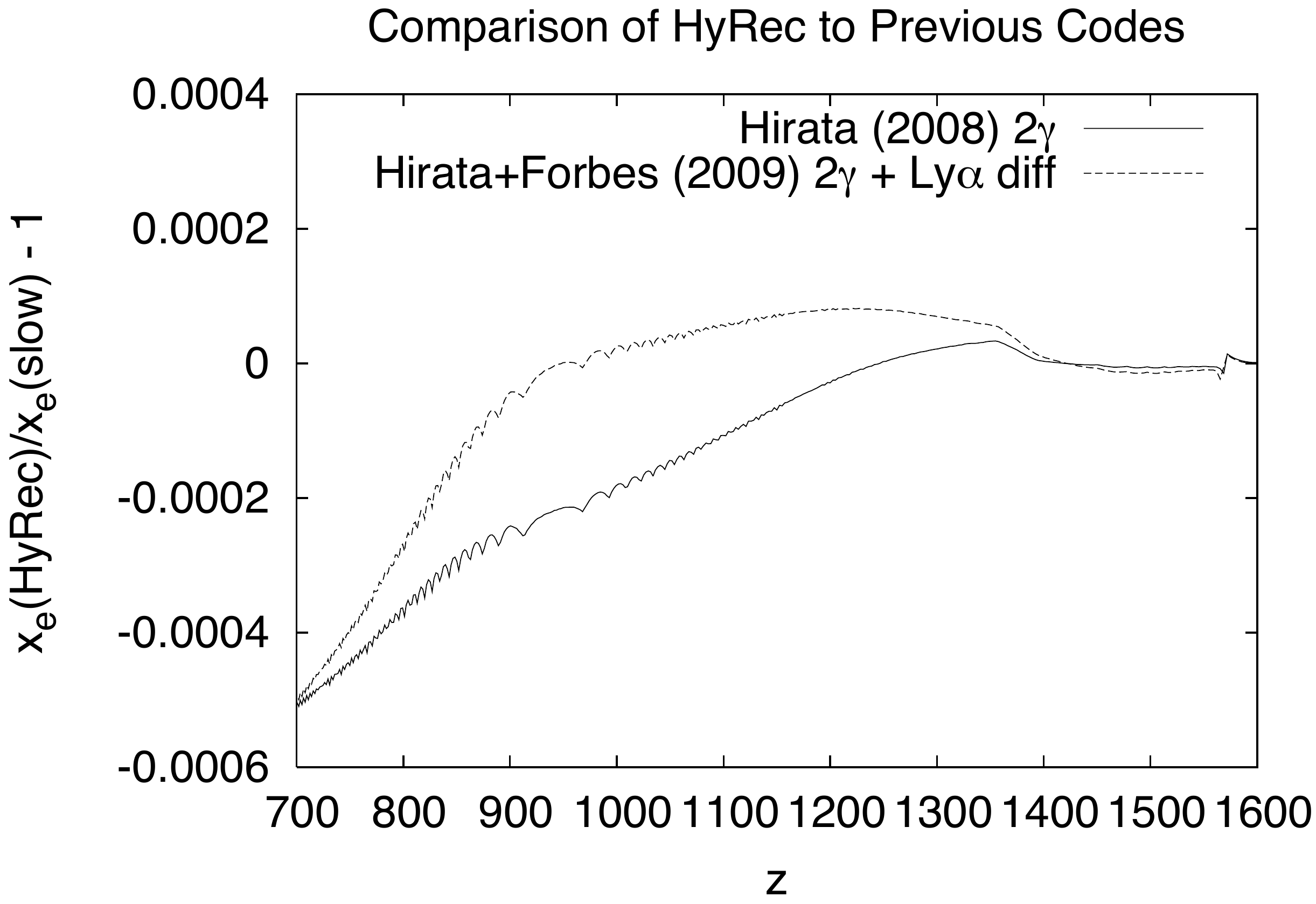}
\caption{Comparison of \textsc{HyRec} with the MLA+two-photon code of Hirata (2008) and to the results of  Hirata \& Forbes (2009) who account for Ly$\alpha$ diffusion. See text for comments.}
\label{fig:hfcompare}
\end{figure}

We compared our results to the two-photon MLA code of Hirata \cite{Hirata_2photon}, as well as to the results of Hirata \& Forbes \cite{Hirata_Forbes} for frequency diffusion. For this comparison, we use $n_{\max} = 30$. The result of the comparison is shown in Fig.~\ref{fig:hfcompare}. The maximum difference between the codes for $700 < z < 1600$ is $|\Delta x_e|/x_e = 0.0005$. The increase of the relative difference at late times is most likely due to small differences in the bound-free rates, which are computed with different methods (we use the recursion relations of Ref.~\cite{Burgess} whereas Hirata (2008) directly integrates the products of wave functions to compute matrix elements). This difference remains even when switching off two-photon processes. The $\sim 3\times 10^{-5}$ kink at $z \sim 1570$ is a startup transient due to switching from the post-Saha solution to solving the full ODE. The kink at $z = 1350$ is due to the Ly$\beta$ photons emitted at $z = 1600$ starting to redshift into Ly$\alpha$. Overall, the agreement is excellent ($|\Delta x_e|/x_e < 10^{-4}$ for $z > 900$), even though the codes use different methods to compute atomic rates and different approaches for solving the MLA problem and treating Ly$\alpha$ frequency diffusion.


\section{Helium recombination} \label{sec:helium}

Roughly 14\%\ of the electrons in the Universe are associated with helium rather than hydrogen, so it is critical to model helium recombination as well 
\cite{Hu1995, Hirata_SwitzerIII, Rubino08_He, Kholupenko08_He}.  Because the ionization energy of helium is greater than that of hydrogen ($E_{I_1} = 24.6$ eV 
for He I and $E_{I_2} = 54.4$ eV for He II), helium recombines earlier than hydrogen, well before the epoch of last scattering.  Therefore, we do not observe 
helium recombination directly: rather it affects the diffusion of photons at early times and hence controls the Silk damping length \cite{Hu1995}.  A 
faster helium recombination (smaller $x_e$) leads to a longer photon mean free path and hence a larger damping length.  The net effect is then to 
reduce the high-$\ell$ multipoles of the CMB temperature and polarization power spectra \cite{Hirata_SwitzerIII}.  However, since most of the Silk 
damping occurs at later times, we do not need extraordinary accuracy in following helium recombination: the corrections identified by 
Refs.~\cite{Hirata_SwitzerI, Hirata_SwitzerII, Hirata_SwitzerIII}, which changed $x_e$ by up to 3\%\ at $z\approx 1800$, amounted to a $\sim 1\sigma$ 
correcton for {\slshape Planck} and $\sim 8\sigma$ for a hypothetical cosmic variance limited experiment to $\ell=3000$.  Therefore, a helium 
recombination code accurate to $\sim 0.3$\%\ should reduce any residual errors to the point of being negligible for {\slshape Planck}.  Here we 
describe our ``fast'' helium recombination code.

Throughout the helium recombination process, we may take a single temperature $T\equiv T_{\rm m}=T_{\rm r}$.  We make several other crude approximations, as detailed below.  While analytically motivated, their quantitative justification rests on the comparison to the ``full'' version of our calculations \cite{Hirata_SwitzerIII}.  For the equilibrium calculations, the He I level energies and are obtained from the NIST database (itself based on the compilation of Ref.~\cite{MWD06}).  The bound-bound Einstein coefficients are obtained from Ref.~\cite{MD07}: $A[2^1P^o-1^1S] = 1.7989\times 10^9$ s$^{-1}$ and $A[2^3P^o-1^1S] = 177.58$ s$^{-1}$.  The bound-free rates are not required here since the excited levels of He I remain in equilibrium with the continuum (i.e. these rates may simply be taken to be ``fast'').

\subsection{He III$\rightarrow$II recombination}

The He III$\rightarrow$II recombination has previously been found to be in Saha equilibrium to high accuracy \cite{Hirata_SwitzerI, Rubino08_He} and is 
followed using the Saha equation: 
\beq
\frac{q(1+f_{\rm He} + q)}{f_{\rm He} - q} = s \equiv g_e \rme^{-E_{I_2}/T},\label{eq:s-heii}
\eeq
where $q \equiv x_{\rm HeIII}$ and we assume that the rest of the helium is singly ionized, $x_{\rm HeII}=f_{\rm He}-q$, and all hydrogen is ionized, $x_e=1+f_{\rm He}+q$.
In Eq.~(\ref{eq:s-heii}), $E_{I_2}$ is the second ionization energy of helium and $g_e$ is given by Eq.~(\ref{eq:ge}) but using the appropriate reduced mass of the electron-He III system. We first solve Eq.~(\ref{eq:s-heii}) for $q$:
\beq
q = \frac{2sf_{\rm He}}{1 + f_{\rm He} + s} \left(1 + \sqrt{1 + \frac{4sf_{\rm He}}{(1 + f_{\rm He} + s)^2}}\right)^{-1}.
\eeq
We then obtain the free electron fraction from $x_e = 1 + f_{\rm He} + q$.


This equation is used to obtain $x_e$ until $q = x_{\rm HeIII} = 10^{-9}$ (which corresponds to $z\sim 4000$).

\subsection{He II$\rightarrow$I recombination: Near-equilibrium stage}
In contrast to He III$\rightarrow$II recombination, the He II$\rightarrow$I recombination is a highly non-equilibrium process (it occurs at much 
lower density and in a weaker radiation field, with a much slower $2\gamma$ decay process, and the excited levels of He I are much closer to the 
continuum as a fraction of the ionization energy than those of He II).  Therefore, it must be followed in several stages.

The first is the near-equilibrium stage, before the main He I resonance line (2$^1P^o$--1$^1S$ at 584 \AA) develops sufficient optical depth to push helium 
recombination out of equilibrium.  In this stage, the free electron fraction is close to the Saha solution $x_e^{\rm Saha} = 1+q$, where $q \equiv x_{\rm HeII}$ satisfies 
\beq
\frac{q(1+q)}{f_{\rm He} - q} = s \equiv 4 g_e \rme^{-E_{I_1}/T},\label{eq:s-hei}
\eeq
and we assume that all hydrogen is ionized and that the helium is distributed between He$^0$ and He$^+$. In Eq.~(\ref{eq:s-hei}), $E_{I_1}$ is the first ionization energy of Helium, and one should use the appropriate reduced mass of the electron-He II system in $g_e$. The additional factor of 4 relative to Eq.~(\ref{eq:s-heii}) is due to the lower spin degeneracy of He I. Again, we first solve Eq.~(\ref{eq:s-hei}) for $q$:
\beq
q = \frac{2 s f_{\rm He}}{1 + s}\left( 1+ \sqrt{1 + \frac{4 s f_{\rm He}}{(1 + s)^2}}\right)^{-1},
\eeq
from which we get $x_e^{\rm Saha} = 1 + q$. We then obtain the post-Saha expansion for the free electron fraction $x_e = x_e^{\rm Saha} + \Delta x_e$ as described in Appendix \ref{app:post-Saha}. We do so until $\Delta x_e $ reaches $5\times 10^{-4}$, which corresponds to $2500 < z < 3000$ depending on cosmology. We checked that using the post-Saha expansion until $\Delta x_e = 10^{-5}$ instead and then numerically integrating the ODE for $x_e$ (described in the next section) leads to maximum changes of $\sim 3 \times 10^{-4}$ in the free electron fraction.  


\subsection{He II$\rightarrow$I recombination: Non-equilibrium stage}

At lower redshifts, one must follow helium recombination carefully, via a Peebles-style ODE \cite{Peebles}, i.e. an equation of the form
\begin{equation}
\dot{x}_e = -{\cal F}(x_e,z;{\bf p}),
\label{eq:ODE-He}
\end{equation}
where ${\bf p}$ is the vector of cosmological parameters.  Despite the complicated radiative transfer physics in the helium problem, a single ODE 
turns out to suffice because the portion of the ultraviolet spectrum that is relevant is relatively narrow and may be treated as in steady state (it 
encompasses the 584 and 591 \AA\ lines). The construction of the ${\cal F}$ function is however more complicated than the Peebles ODE for hydrogen.

The excited levels of He I ($n\ge 2$) have been found to 
remain in Saha equilibrium with the continuum throughout the process since the strong CMB blackbody ionizes them far faster than 
they can reach the ground state \cite{Hirata_SwitzerI}.  The significant processes for net decays to the ground state are four-fold:
\begin{itemize}
\item The two-photon process, He(2$^1S$)$\rightarrow$He(1$^1S)+\gamma+\gamma$.
\item Emission of photons via the main resonance line 2$^1P^o$--1$^1S$ with $\lambda=584$ \AA, followed by redshifting out of the line (the He I 
analogue of the H I Lyman-$\alpha$ escape process).
\item The absorption of He I 584 \AA\ photons by H(1s) atoms via photoionization, H(1s)+$\gamma\rightarrow$H$^++e^-$.  The electron rapidly thermalizes 
its energy, leading to a loss of a resonance line photon and a net decay of He I.
\item Emission of photons in the intercombination line He I] 2$^3P^o$--1$^1S$ with $\lambda=591$ \AA.  This line has Sobolev optical depth of order 
unity during He I recombination so the full Sobolev escape probability formula must be used.
\end{itemize}

In constructing the function ${\cal F}$ above, we take several steps.  First, we compute the abundance of the species H I, H II, He I and He II.  The formulae 
for the latter are $x_{\rm HeII} = x_e - 1 + x_{\rm HI}$ and $x_{\rm HeI} = f_{\rm He}-x_{\rm HeII}$, but these require knowledge of the H I fraction.  
This cannot be completely ignored, but it is small and so the Saha equation for H (Eq.~\ref{eq:Hsaha}) suffices for this purpose.\footnote{Eq.~(\ref{eq:Hsaha}) technically 
neglects the contribution of helium to the electron abundance, but this correction is small in the very latest stages of He recombination when the H 
I correction becomes significant.}

Next is the determination of the excited level (2$^1S$ and 2$^1P$) abundances assuming equilibrium with the continuum,
\beq
x_{[2^1S]} = \frac{1}{4 g_e}\rme^{-E_{[2^1S]}/T} x_e (x_e - 1)
\eeq
and 
$x_{[2^1P]} = 3 x_{[2^1S]} \exp\left[-(E_{[2^1P]} - E_{[2^1S]})/T\right]$.
 
Armed with this information, we proceed to investigate each decay mechanism.

\subsubsection{The two-photon process}

The downward rate for the He I 2$\gamma$ decay is $\Lambda = $50.94 s$^{-1}$ \cite{Drake86}, so we find a downward decay rate
\beq
{\cal F}^{(2\gamma)}(x_e,z;{\bf p}) = \Lambda\left(x_{[2^1S]} - \rme^{-(E_{[2^1S]}-E_{[1^1S]})/T} x_{\rm He I}\right).
\eeq

\subsubsection{The 584 \AA\ line}

The rate of emission of photons in the 584 \AA\ line is complicated because photons in this line experience multiple processes: (i) ``true'' absorption 
and emission, (ii) resonant scattering (which redistributes photons in frequency since in the comoving frame the ingoing and outgoing photons need not have 
the same frequency), (iii) H I photoionization opacity (which has absorption and emission), and (iv) redshifting due to Hubble expansion.  We construct 
here a highly simplified model of these processes; the ``full'' treatment can be found in Ref.~\cite{Hirata_SwitzerI}.  Ref.~\cite{Hirata_SwitzerI} 
also showed that the resonant scatterings can be neglected.  It should be noted that in the case of the 584 \AA\ line, the nontrivial physics takes place in the damping wings: the line center is very optically thick, and the reaction
\begin{equation}
{\rm He}(2^1P^o) \leftrightarrow {\rm He}(1^1S) + \gamma
\end{equation}
reaches equilibrium there (analogous to the H I Lyman-$\alpha$ line).

In this situation, the radiative transfer equation for the photon occupation number $f_\nu$ in the vicinity of the 584 \AA\ line can be simplified to
[Eq.~(36) of Ref.~\cite{Hirata_SwitzerI}, with individual expressions substituted in]
\begin{eqnarray}
\dot f_\nu &=&  H\nu\frac{\partial f_\nu}{\partial\nu} + H\nu\eta_{\rm c}(e^{-h\nu/T} - f_\nu)
\nonumber \\ &&
+ H\nu\tau_{\rm abs}\phi(\nu) \left( f_{584}^{\rm eq} - f_\nu \right),
\label{eq:fnu-He}
\end{eqnarray}
where $\eta_{\rm c}$ is the H I opacity in units of absorption optical depth per unit frequency (i.e. $H\nu\eta_{\rm c}$ is the absorption optical 
depth per unit time), and $f_{584}^{\rm eq} \equiv x_{[2^1P^o]}/(3x_{[1^1S]})$ is the equilibrium photon occupation number in the 584 \AA\ line. In Eq.~(\ref{eq:fnu-He}), the $e^{-h\nu/T}$ term corresponds to emission from direct H I recombinations to the ground state (i.e. the inverse process of photoionization); $\tau_{\rm abs}$ is the Sobolev optical depth to true absorption in the 584 \AA\ line; $\phi(\nu)$ is the 584 \AA\ line profile normalized by $\int\phi(\nu)\,d\nu=1$. We neglect stimulated emission processes at 584 \AA, since the photon phase space density is $\ll 1$. In steady state, the left-hand side of Eq.~(\ref{eq:fnu-He}) is approximated as
zero. The remaining contributions are as follows.

The H I continuum optical depth is, assuming an H I abundance in Saha equilibrium with H II 
(assumed to be nearly all hydrogen, i.e. $x_p\approx 1$),
\begin{equation}
\eta_{\rm c} = \frac{\rme^{E_I/T}}{g_e} \frac{\sigma_{\rm pi} c n_{\rm H}x_e}{H\nu}.
\end{equation}
[This comes from combining Eqs.~(27) and (28) of Ref.~\cite{Hirata_SwitzerI}.]
Here $\sigma_{\rm pi}$ is the photoionization cross section of H I at $\lambda = 584$ \AA\ and $E_I$ is the hydrogen ionization energy.

The line profile is in principle a Voigt profile.  However near line center where $\phi(\nu)$ becomes large, we have $f_\nu \rightarrow 
f_{584}^{\rm eq}$ irrespective of the details in order to keep Eq.~(\ref{eq:fnu-He}) finite.  Therefore, we approximate it by the damping wing 
approximation,
\begin{equation}
\phi(\nu) \approx \frac{\Gamma}{4\pi^2(\nu-\nu_0)^2},
\end{equation}
where $\nu_0$ is the central frequency of the line and $\Gamma$ is the intrinsic width.  The latter is the sum of the rates for all processes that 
depopulate $2^1P^o$ (the width of the $1^1S$ state is negligible), i.e. we may write $\Gamma = A_{584} + \Gamma_{\rm other}$, where $A_{584}$ is the 
contribution from $2^1P^o\rightarrow 1^1S$ and $\Gamma_{\rm other}$ is the contribution from all other states:
\begin{eqnarray}
\Gamma_{\rm other} &=&  \frac{A_{[2^1P^o], [2^1S]}}{1-\rme^{-(E_{[2^1P^o]}-E_{[2^1S]})/T}}
\nonumber \\ &&
 + \sum_i \frac{g_i}3 \frac{A_{i, [2^1P^o]}}{\rme^{(E_i-E_{[2^1P^o]})/T}-1},
\end{eqnarray}
where the first term corresponds to decays to $2^11S$ (supplemented by stimulated transitions) and the second term corresponds to absorptions from 
$2^1P^o$ to a higher level $i$.  We include the levels $n^1S$ and $n^1D$ for $3\le n\le 5$.  In principle we should include higher $n$ and the 
continuum levels, but in practice the first few levels dominate the sum because of the exponential factor.

We next need the optical depth to true absorption, which is the Sobolev optical depth $\tau_{\rm S}$ times the fraction of photon absorptions 
$1^1S\rightarrow 
2^1P^o$ that are true absorptions (i.e. do not immediately decay back to $1^1S$, but rather visit another level).  This fraction is $\Gamma_{\rm 
other}/\Gamma$.  Thus
\begin{equation}
\tau_{\rm abs}\phi(\nu) = \frac{\Gamma_{\rm other}}{\Gamma} \tau_{\rm S} \frac{\Gamma}{4\pi^2(\nu-\nu_0)^2}
=  \frac{\tau_{\rm S} \Gamma_{\rm other}}{4\pi^2(\nu-\nu_0)^2}.
\end{equation}

With these results, and neglecting the variation of the blackbody function $e^{-h\nu/T}$ across the line, Eq.~(\ref{eq:fnu-He}) simplifies to
\begin{equation}
\frac{\partial f_\nu}{\partial\nu} = \eta_{\rm c}(f_\nu - e^{-h\nu_0/T}) + \frac{ \tau_{\rm S}\Gamma_{\rm other}}{4\pi^2(\nu-\nu_0)^2}
\left( f_\nu - f_{584}^{\rm eq}\right).
\end{equation}
The next simplification of this equation occurs if we re-scale both the frequency and the phase space density axes as
\begin{equation}
y = 4\pi^2 \frac{\nu-\nu_0}{\tau_{\rm S}\Gamma_{\rm other}}
\end{equation}
and
\begin{equation}
\xi(y) = \frac{ f_\nu - e^{-h\nu_0/T} }{ f_{584}^{\rm eq} - e^{-h\nu_0/T} },
\end{equation}
leading to
\begin{equation}
\frac{d\xi}{dy} = \tau_{\rm c} \xi + \frac{\xi-1}{y^2},
\label{eq:xidifeq}
\end{equation}
where
\beq
\tau_{\rm c} \equiv \frac{\tau_{\rm S}\Gamma_{\rm other}\eta_{\rm c}}{4\pi^2}.
\eeq
For blackbody radiation entering the line, we have the boundary condition $\xi(+\infty)=0$.

This reduces the radiative transfer equation to a single ODE that depends on the single dimensionless parameter $\tau_{\rm c}$, which is (roughly speaking) the optical depth to H I photoionization within the part of the line that is optically thick 
to true absorption by He I 584 \AA.  This parameter is exponentially increasing in the early parts of helium recombination, and becomes of order unity 
at $z\approx 2100$.  It is the parameter that controls which process is a more important sink for resonance line photons: H I continuum opacity 
($\tau_{\rm c}\gg1$) or escape via redshifting ($\tau_{\rm c}\ll1$).

The net rate at which photons are emitted in the He I line in photons per H nucleus per unit time is then obtained by integrating the 
absorption/emission term in the rate equation,
\beq
{\cal F}^{(584)} = \int \frac{8\pi\nu^2}{n_{\rm H}c^3} 
H\nu\tau_{\rm abs}\phi(\nu) \left( f_{584}^{\rm eq} - f_\nu \right)\, d\nu.
\eeq
This integral can be converted into an integral over $y$ and $f_\nu$ can be replaced with $\xi(y)$; making these substitutions and approximating $\nu\approx \nu_0$ in the prefactors gives
\begin{equation}
{\cal F}^{(584)} = \frac{8\pi H\nu_0^3}{n_{\rm H}c^3}
\left( f_{584}^{\rm eq}- e^{-h\nu_0/T} \right) {\cal E},
\label{eq:F-He}
\end{equation}
where
\begin{equation}
{\cal E} = \int \frac{1-\xi(y)}{y^2} dy
\label{eq:E-He}
\end{equation}
depends on the single parameter $\tau_{\rm c}$.  We note that without the factor of ${\cal E}$, Eq.~(\ref{eq:F-He}) would be the standard decay rate formula with escape probability $P_{\rm esc}=1/\tau_{\rm S}$ (appropriate in the high optical depth limit).  Thus ${\cal E}$ can be thought of as a correction factor associated with the H I continuum opacity.

\begin{figure}
\includegraphics[width = 85 mm]{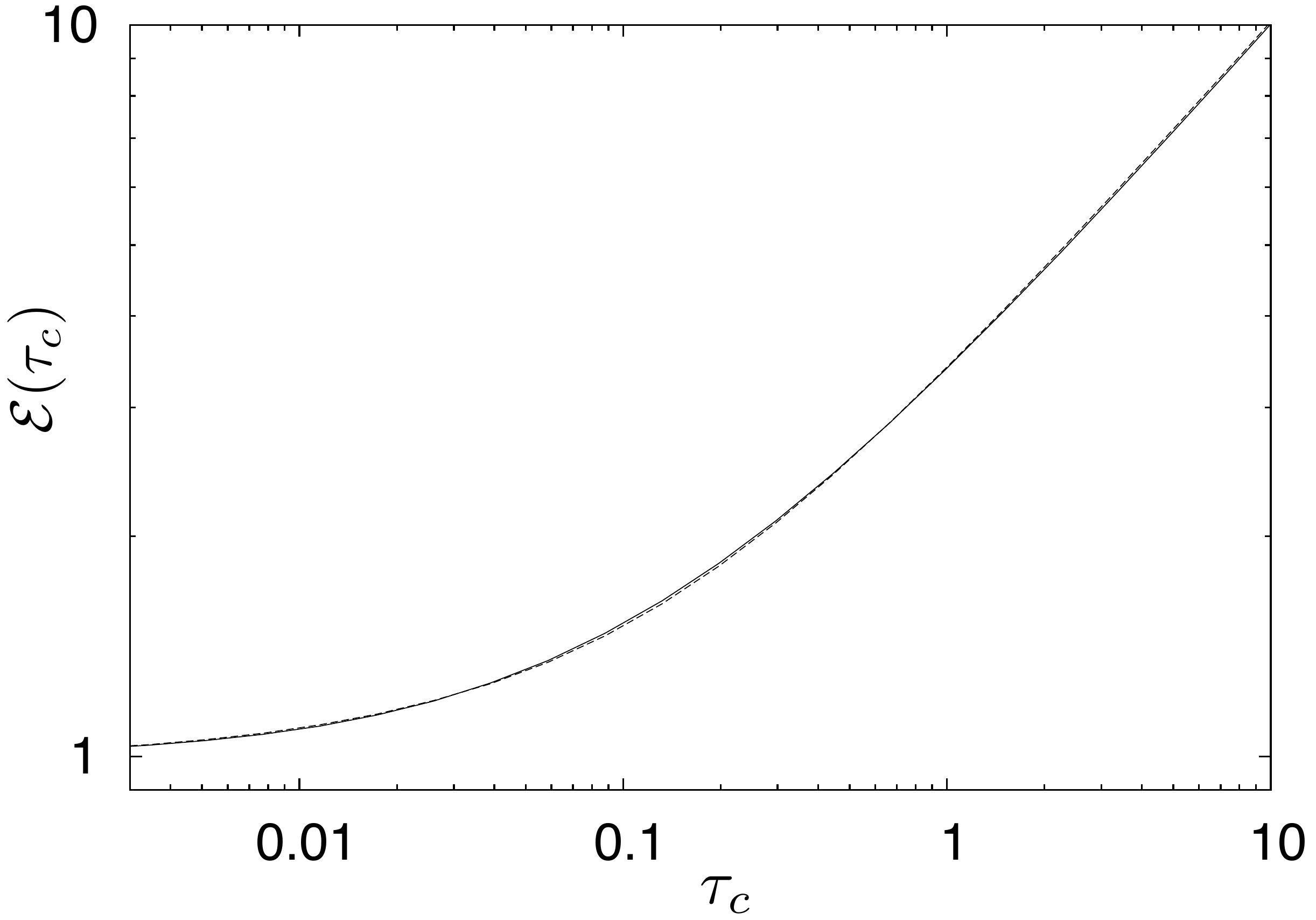}
\caption{\label{fig:E}The function ${\cal E}(\tau_{\rm c})$ (solid line) and our approximation (dashed line).}
\end{figure}

Unfortunately, the integral in Eq.~(\ref{eq:E-He}) is not amenable to direct calculation, since the integrand is ill-conditioned near $y\approx 0$.  We may obtain an alternative form by taking the integral from $y=-Y$ to $y=+Y$ (we will take the limit $Y\rightarrow\infty$), plugging in Eq.~(\ref{eq:xidifeq}) into the integrand, and then realizing that $\xi(+Y)\rightarrow 0$ due to our boundary condition, we find
\begin{equation}
{\cal E} = \xi(-Y) + \tau_{\rm c} \int_{-Y}^Y \xi(y)\,dy.
\end{equation}
Formally, for $\tau_{\rm c}>0$, we have $\xi(-Y)\rightarrow 0$ for $Y\rightarrow \infty$, but numerically one must keep the first term at small $\tau_{\rm c}$.  Equation~(\ref{eq:xidifeq}) is stiff and can be solved by the backward Euler method; however even this is too slow to use in a ``fast'' recombination code.  Therefore, we have constructed a fitting function for ${\cal E}$, valid to within 0.8\%\ for all positive $\tau_{\rm c}$:
\begin{equation}
{\cal E}(\tau_{\rm c}) \approx \sqrt{1+\pi^2\tau_{\rm c}} + \frac{7.74\tau_{\rm c}}{1+70\tau_{\rm c}}.
\label{eq:E-approx}
\end{equation}
Note that this has the correct limiting behavior ${\cal E}\rightarrow 1$ for $\tau_{\rm c}\rightarrow 0$, and for positive $\tau_{\rm c}$ has ${\cal E}>1$, as one would expect.  The full function ${\cal E}(\tau_{\rm c})$ and the approximation of Eq.~(\ref{eq:E-approx}) are shown in Fig.~\ref{fig:E}.

The use of Eqs.~(\ref{eq:F-He}) and (\ref{eq:E-approx}) are sufficient to solve for the behavior of the 584 \AA\ line.

\subsubsection{The 591 \AA\ line}

The intercombination line He I] 2$^3P^o_1\rightarrow 1^1S$ at 591 \AA\ can reach optical depths of order unity during helium recombination.  Therefore it must be considered carefully.  However, because the damping wings are optically thin, it is only the line center that is of interest.  Because there is negligible H I continuum opacity in the core during helium recombination, we use the Sobolev approximation for the 591 \AA\ line.  If there were no 584 \AA\ line, then we would have blackbody radiation entering the line and could write the net decay rate as
\begin{equation}
\frac{8\pi H\nu_0^3}{n_{\rm H}c^3} (1-e^{-\tau_{591}})
\left( f_{591}^{\rm eq} - e^{-(E_{[2^3P^o]}-E_{[1^1S]})/T} \right),
\label{eq:591}
\end{equation}
where $\tau_{591}$ is the optical depth of the 591 \AA\ line, and $f_{591}^{\rm eq} \equiv x_{[2^3P^o]}/(3x_{[1^1S]})$ is the equilibrium value of the photon occupation number in the line. Based on the ratio of Einstein coefficients \cite{MD07}, we take $\tau_{591}=1.023\times 10^{-7}\tau_{584}$.

However, the 591 \AA\ line can also absorb photons that redshifted out of the 584 \AA\ line.  We treat this problem by an ``on-the-spot'' approximation: we assume that for each photon emitted in the 584 \AA\ line, there is a probability that the photon is re-absorbed of
\begin{equation}
P_{\rm reabs} = e^{-\eta_{\rm c}(E_{[2^1P^o]}-E_{[2^3P^o]})/h} (1-e^{-\tau_{591}}),
\end{equation}
where the first factor is the probability that the photon can redshift from 584 \AA\ to 591 \AA\ without being destroyed by H I continuum opacity, and the second factor is the probability that, once it reaches the 591 \AA\ line, the photon is indeed absorbed.  Thus the 591 \AA\ line has two effects: first, it contributes to the net formation of the He I ground state in accordance with Eq.~(\ref{eq:591}), and second it reduces the net formation from the 584 \AA\ line by a factor of $1-P_{\rm reabs}$.  Thus we write
\begin{eqnarray}
{\cal F}^{(591)} &=& 
\frac{8\pi H\nu_0^3}{n_{\rm H}c^3} (1-e^{-\tau_{591}})
\nonumber \\ && \times
\left( f_{591}^{\rm eq} - e^{-(E_{[2^3P^o]}-E_{[1^1S]})/T} \right)
\nonumber \\ && 
 - P_{\rm reabs}{\cal F}^{(584)}.
\end{eqnarray}

\subsubsection{Hydrogen recombination corrections}

As a final step, we include in our rate equation the contribution to the ionization fraction due to changes in the hydrogen Saha equilibrium:
\begin{equation}
{\cal F}^{\rm(H)} = -\left.\frac{dx_{\rm HI}}{dt}\right|_{\rm Saha},
\end{equation}
which is obtained by two-sided finite differencing with $\Delta z=\pm0.5$.  This contribution to $\dot x_e$ is necessary for full accuracy at the very last stages of He recombination.

\subsubsection{Integration details}

We may now construct the overall electron fraction evolution, ${\cal F}={\cal F}^{(2\gamma)} + {\cal F}^{(584)} + {\cal F}^{(591)} + {\cal F}^{\rm(H)}$.
The equation is stiff at early times, which is why we do not turn it on until the free electron fraction given by the post-Saha expansion differs from the Saha equilibrium value by more than $5\times 10^{-4}$. 
We then use the ODE integrator described in Appendix \ref{app:ODE}.

The treatment of helium is turned off once the abundance of He$^+$ ions is less than $10^{-4}$ per hydrogen atom, after which we switch to following hydrogen recombination (we checked that hydrogen and helium recombination never overlap and can be followed separateley).

\subsection{Comparison to detailed calculations}

\begin{figure}
\includegraphics[width = 85 mm]{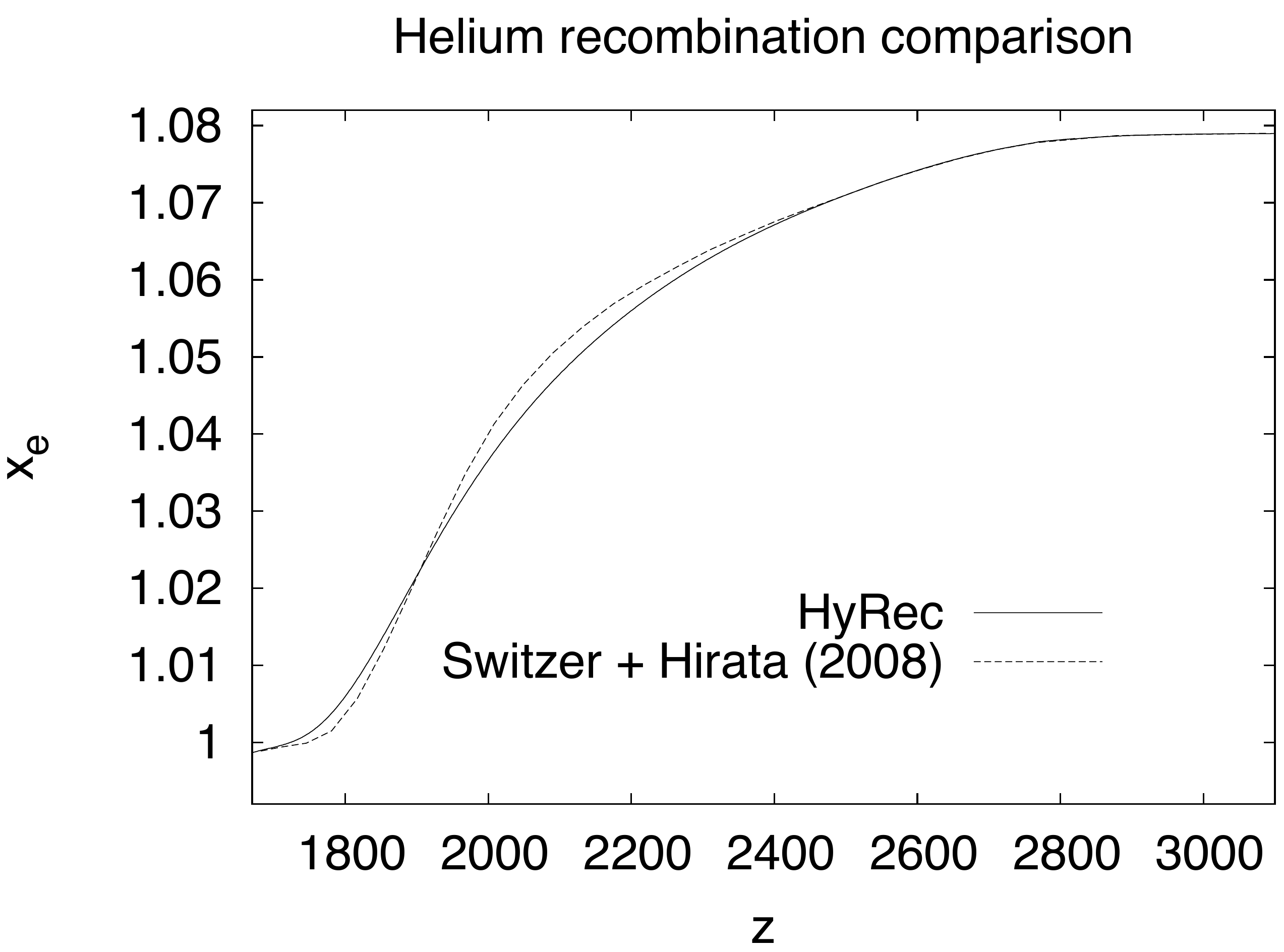}
\caption{\label{fig:shcompare}A comparison of the {\sc HyRec} He II$\rightarrow$I recombination history to that computed by the ``full physics'' code of Switzer \& Hirata \cite{Hirata_SwitzerIII}.  The maximum deviation is 0.3\%.}
\end{figure}

The ultimate test of the sweeping approximations made in this section is their comparison against more detailed computations of the He I recombination history.  We compare against Switzer \& Hirata \cite{Hirata_SwitzerIII} in Fig.~\ref{fig:shcompare}.  The maximum error is 0.3\%, which is roughly equal to the stated theoretical uncertainty in the Switzer \& Hirata calculation.

\section{Conclusions} \label{sec:conclusion}

We have presented a complete treatment of primordial hydrogen and helium recombination, including all the effects that have been shown to be important so far. Our computation accounts for the multilevel character of hydrogen and the non-equilibrium of angular momentum substates, radiative feedbacks, two-photon transitions, and frequency diffusion in Ly$\alpha$ for hydrogen recombination. For helium recombination, we account for HI continuum opacity in the He I $2^1P^o-1^1S$ line, decays in the $2^3P^o-1^1S$ intercombination line, and feedback between these lines.
We have implemented all these effects in a single recombination code, \textsc{HyRec}, which can compute a recombination history in $\sim 2$ seconds on a standard laptop for a given set of cosmological parameters. Provided collisional transitions can be neglected (which remains to be established), we estimate the errors of our computation to be a few times $10^{-3}$ during helium recombination and a few times $10^{-4}$ during hydrogen recombination, including both numerical errors and errors due to the assumptions and approximations made for physical effects. If collisional transitions are shown to have a significant effect on recombination, our code can be easily updated to account for them with very little loss of computational efficiency.

It has been argued that corrections to the recombination history due to radiative transfer effects are relatively independent of cosmology \cite{Rubino10}, and that one could therefore compute them once and use the resulting correction function to account for them for any given cosmology. Alternatively, one might run a grid of recombination histories for different cosmologies and construct a fitting function \cite{Lewis06, RICO}.  Our point of view here is that the physics of primordial recombination is simple enough, and an exact calculation from first principles is now fast enough that there should be no reason to use fudge factors and approximate correction functions.  This is especially relevant if one wishes to extend the standard recombination calculation by introducing ``exotic'' new physics. We would like to emphasize that the fast computation presented here, using the EMLA method, is very well adapted for the computation of the recombination history, but that the standard MLA approach and fast interpolation methods may still be useful for the computation of the recombination spectrum. 

We believe our code is accurate enough (aside from neglecting collisional transitions) and has a sufficiently small runtime to be incorporated in Monte Carlo Markov chains for upcoming CMB data analysis from the \emph{Planck} mission.

\section*{Acknowledgments}
The authors thank Jens Chluba for stimulating conversations on the physics of recombination and Daniel Grin for carefully reading the manuscript of this paper. Y. A-H. and C. H. are supported by the U.S. Department of Energy (DE-FG03-92-ER40701) and the National Science Foundation (AST-0807337). C. H. is supported by the Alfred P. Sloan Foundation and the David \& Lucile Packard Foundation.

\appendix

\section{Proof of some relations involving effective rates}\label{app:effective-rates}

\subsection{Preliminaries}

Here we use the same notation as in Paper I. Capital indices $K,L$ refer to ``interior'' excited states, and lower-case indices $i,j$ refer to ``interface'' excited states. We define the rate matrix $\mathbf{M}$ whith coefficients:
\beq
M_{KL} = \delta_{KL} \Gamma_K - (1 - \delta_{KL}) R_{K,L}.
\eeq
In Paper I, we have shown that the populations of the interface states are given by:
\beq
X_K = \sum_L \left(\mathbf{M}^{-1}\right)_{LK}\left[n_{\rm H} x_e^2 \alpha_L + \sum_i x_i R_{i,L}\right].\label{eq:XK-sol}
\eeq
We also showed that the probabilities $P_K^i$ and $P_K^e$ are given by
\beq
P_K^i = \sum_L \left(\mathbf{M}^{-1}\right)_{KL} R_{L,i}, \label{eq:PKi-sol}
\eeq
and
\beq
P_K^e = \sum_L \left(\mathbf{M}^{-1}\right)_{KL} \beta_L. \label{eq:PKe-sol}
\eeq
In Appendix C of Paper I, we showed that $\mathbf{M}$ satisfies the following detailed balance relation:
\beq
Q_K \left(\mathbf{M}^{-1}\right)_{KL} = Q_L \left(\mathbf{M}^{-1}\right)_{LK}, \label{eq:MKL-db}
\eeq
where $Q_K = g_K \rme^{-E_K/\Tr}$ is the contribution of individual states to the partition function and $g_K$ is the degeneracy of the state $K$.

\subsection{Rate of change of the free electron fraction} \label{app:xedot}

In this section we derive Eq.~(\ref{eq:xedot EMLA}), which was not derived in Paper I. In the standard MLA formulation, the rate of change of the free electron fraction can be written as:
\barr
\dot{x}_e = &-& \sum_{K} \left[n_{\rm H} x_e^2 \alpha_K - X_K \beta_K\right]\nonumber\\
                  &-& \sum_{i}\left[n_{\rm H} x_e^2 \alpha_i - x_i \beta_i\right]. \label{eq:xedot formal}
\earr
This formula is never used in standard MLA codes, as it requires a summation over a large number of nearly cancelling terms, and MLA codes use instead $\dot{x}_e = - \dot{x}_{1s}$ to compute the rate of change of the free electron fraction. Eq.~(\ref{eq:xedot formal}) remains however formally correct. Using Eqs.~(\ref{eq:XK-sol}) and (\ref{eq:PKe-sol}), we rewrite:
\barr
\sum_{K} X_K\beta_K &=&  \sum_{K,L}  \beta_K \left(\mathbf{M}^{-1}\right)_{LK}\left[n_{\rm H} x_e^2 \alpha_L + \sum_i x_i R_{i,L}\right]\nonumber\\
&=& \sum_L P_L^e \left[n_{\rm H} x_e^2 \alpha_L + \sum_i x_i R_{i,L}\right]\nonumber\\
&=& \sum_L n_{\rm H} x_e^2 \alpha_L - n_{\rm H} x_e^2 \sum_i \sum_L \alpha_L P_L^i \nonumber\\
&&+ \sum_i x_i \sum_L R_{i,L}P_L^e,
\earr
where in the last equality we have used the complementarity relation $\sum_i P_K^i + P_K^e = 1$. Inserting this result into Eq.~(\ref{eq:xedot formal}), and using the definitions of the effective recombination coefficients and photoionization rates Eqs.~(\ref{eq:Ai}) and (\ref{eq:Bi}), we immediately recover Eq.~(\ref{eq:xedot EMLA}).

\subsection{Proof of Eq.~(\ref{eq:Ri1s-new})} \label{app:proof Ri1s}

Consider Eq.~(\ref{eq:tildePK1s}) with $\tilde{\Gamma}_K \approx \Gamma_K(\Tr)$. The formal solution for the $\tilde{P}_K^{1s}$ is given by:
\beq
\tilde{P}_K^{1s} = \sum_{L} \left(\mathbf{M}^{-1}\right)_{KL} \tilde{R}_{L,1s}. 
\eeq
Therefore one may rewrite Eq.~(\ref{eq:Ri1s}), for $i = 2s, 2p$: 
\beq
\tilde{\mathcal{R}}_{i,1s} = \tilde{R}_{i,1s} + \sum_{K} \lambda_{i,K}(\Tr) \tilde{R}_{K,1s},
\eeq 
where we have defined
\barr
\lambda_{i,K}(\Tr) &\equiv& \sum_{L} R_{i,L}\left(\mathbf{M}^{-1}\right)_{LK} \nonumber\\
&=& \sum_{L} R_{i,L}\frac{Q_K}{Q_L}\left(\mathbf{M}^{-1}\right)_{KL} \nonumber\\
&=&\sum_L R_{L,i} \frac{Q_K}{Q_i}\left(\mathbf{M}^{-1}\right)_{KL} \nonumber\\
&=& \frac{g_K}{g_i} \rme^{-E_{Ki}/\Tr}  P_K^{i}(\Tr),
\earr
where in the second line we used Eq.~(\ref{eq:MKL-db}), in the third line we used the detailed balance relation verified by $R_{i,L}$ and $R_{L,i}$, and in the last line we used the formal solution for $P_K^i$, Eq.~(\ref{eq:PKi-sol}). We therefore obtain Eq.~(\ref{eq:Ri1s-new}).

\subsection{Expression of $X_K$ in terms of $x_i, x_e$ and effective rates}\label{app:XK}
Taking $\Tm = \Tr$ and using the detailed balance relation $g_e n_{\rm H} \alpha_L = Q_L \beta_L$, we rewrite Eq.~(\ref{eq:XK-sol}) as follows:
\barr
X_K &=& g_e^{-1} x_e^2  \sum_L Q_L\left(\mathbf{M}^{-1}\right)_{LK} \beta_L \nonumber\\
&&+ \sum_i x_i \sum_L \left(\mathbf{M}^{-1}\right)_{LK} \frac{Q_L}{Q_i} R_{L,i}\nonumber\\
&=& g_e^{-1} x_e^2 Q_K \sum_L \left(\mathbf{M}^{-1}\right)_{KL} \beta_L \nonumber\\
&&+ \sum_i x_i \frac{Q_K}{Q_i} \sum_L \left(\mathbf{M}^{-1}\right)_{KL} R_{L,i},
\earr
where in the last equality we have used Eq.~(\ref{eq:MKL-db}). Using the formal solutions for the probabilities Eqs.~(\ref{eq:PKi-sol}), (\ref{eq:PKe-sol}), we see that we recover Eq.~(\ref{eq:XK-effective}).

\section{Extrapolation of the effective rates to $n_{\max} = \infty$} \label{app:extrapolation}

We have tabulated the effective rates $\mathcal{A}_{2s}(\Tm, \Tr), \mathcal{A}_{2p}(\Tm, \Tr)$ and $\mathcal{R}_{2s,2p}(\Tr)$, including all excited states up to the principal quantum number $n_{\max}$, for several values of $n_{\max}$ up to 600, over the temperature range $0.004~\textrm{eV} \leq \Tr \leq 0.4~\textrm{eV}$, $0.1\leq \Tm/\Tr \leq 1$. This range of temperatures corresponds to $20< z < 1650$ for a wide range of cosmologies. 
For every pair $(\Tm, \Tr)$, we have fitted the effective rates by the following functional form:
\beq
\mathcal{A}_{i}(\Tm, \Tr; n_{\max}) = \mathcal{A}_{i}(\Tm, \Tr; \infty) \left(1 - \frac{\kappa}{(n_{\max})^{\gamma}}\right),
\eeq
and similarly for $\mathcal{R}_{2s,2p}$, where $\kappa$ and $\gamma$ depend on $\Tm$ and $\Tr$ as well as on the coefficient being fitted. This allows us to extrapolate the effective rates to $n_{\max} \rightarrow \infty$. Of course, this is only a formal extrapolation, as for $n$ larger than a few thousands, the excited states of hydrogen are no more well defined (see Ref.~\cite{grinthesis} for a discussion). The extrapolated rates are still more accurate than those computed with a a finite number of states. The residuals of the fit have a maximum relative amplitude of $5 \times 10^{-4}$ over the whole range of temperature considered, for $200 \leq n_{\max} \leq 600$, and more than an order of magnitude smaller on the restricted range $\Tr \geq 0.04$ eV, $\Tm/\Tr \geq 0.8 $ which corresponds to $z \gtrsim 200$ (note that neglecting the overlap of the high-lying Lyman lines leads to errors in the effective rates of similar amplitude \cite{AGH10}). For reference, the maximum relative difference between the effective rates computed with $n_{\max} = 600$ and their extrapolation at $n_{\max} = \infty$ is 0.05 over the whole range of temperature considered, and 0.002 over the restricted range corresponding to $z \gtrsim 200$. We checked that our method recovers the correct case-B recombination coefficient $\alpha_B(\Tm) \equiv \sum_{i=2s,2p}\mathcal{A}_{i}(\Tm, \Tr = 0; \infty)$. Our extrapolated $\alpha_B$ agrees with the fit of Ref.~\cite{alphaB} to better that 0.2 \% for $\Tm > 40$ K, which is the accuracy claimed by the authors of Ref.~\cite{alphaB}. \\
 
\section{Numerical ODE integrator} \label{app:ODE}

For the sake of computational efficiency, we use a second order ODE integrator that uses derivatives computed at previous timesteps. This allows us to evaluate derivatives only once at each timestep. Explicitly, to numerically solve the equation $y'(x) = f(x, y)$, we use evenly spaced steps $\Delta x$, and obtain the solution at the $(i+1)$th step as follows:
\barr
y_{i+1} &=& y_i + \Delta y_i, \nonumber \\
\Delta y_i &=&  \Delta x\left[1.25 y'_i - 0.25 y'_{i-2}\right],
\earr 
where $y'_i = f(x_i, y_i)$ is stored at each timestep for later use. For the case of interest, we have $x = \ln a$, $y = x_e$ and $f = \dot{x}_e/H$.

\section{Post-Saha expansion at early phases of hydrogen and He II$\rightarrow$I recombinations} \label{app:post-Saha}

As explained in Appendix D of Paper I, the ODE describing hydrogen recombination is stiff at $z \gtrsim 1500$ and so is the ODE describing He II$\rightarrow$I recombination at $z\gtrsim 2800$. We therefore use an expansion around the Saha equilibrium solution:
\beq
x_e \approx x_e^{\rm S} + \frac{d (x_e^{\rm S})}{dt} \Bigg{/}\frac{\partial \dot{x}_e}{\partial x_e}\Big{|}_{x_e^{\rm S}}, \label{eq:post-saha}
\eeq
where $x_e^{\rm S}$ is the Saha equilibrium value of the free electron fraction.
\subsection{Hydrogen recombination}
The Saha equilibrium value of the free electron fraction is the solution of the following equation:
\beq
\frac{(x_e^{\rm S})^2}{1-x_e^{\rm S}} = s \equiv g_e \rme^{-E_I/T}, \label{eq:Hsaha}
\eeq
where $g_e$ was given in Eq.~(\ref{eq:ge}) and $T = \Tm = \Tr$ at early times. The numerator in Eq.~(\ref{eq:post-saha}) can be obtained analytically by differentiating Eq.~(\ref{eq:Hsaha}):
\beq
\frac{d (x_e^{\rm S})}{dt} = - \frac{H(\frac{E_I}{T} - \frac32) (x_e^{\rm S})^2}{2 x_e^{\rm S} + s}.
\eeq
For the denominator in Eq.~(\ref{eq:post-saha}), we numerically differentiate the derivative $\dot{x}_e$ obtained when accounting for two-photon processes and diffusion, using a two-sided finite difference with $\Delta x_e =  \pm 0.01(1 - x_e^{\rm S})$.

\subsection{He II$\rightarrow$I recombination}

In that case the free electron fraction in Saha equilibrium is given by $x_e^{\rm S} = 1+ q$, where $q$ can be obtained from Eq.~(\ref{eq:s-hei}). As in the hydrogen case, differentiation of Eq.~(\ref{eq:s-hei}) gives us an analytic expression for the numerator in Eq.~(\ref{eq:post-saha}). We numerically differentiate the derivative $\dot{x}_e$ given by Eq.(\ref{eq:ODE-He}) using a two-sided finite difference with $\Delta x_e =  \pm 0.01(1 + f_{\rm He}- x_e^{\rm S})$.


\bibliography{references}

\end{document}